\documentclass[11pt]{article}
\usepackage[final]{graphics}

\def\line#1{\hbox to \textwidth{#1}}
\catcode`\@=11

\def\thebibliography#1{\section*{REFERENCES}\list{\arabic{enumi}.}
  {\settowidth\labelwidth{#1.}\leftmargin=1.67em
   \labelsep\leftmargin \advance\labelsep-\labelwidth
   \itemsep\z@ \parsep\z@
   \usecounter{enumi}}\def\makelabel##1{\rlap{##1}\hss}%
   \def\newblock{\hskip 0.11em plus 0.33em minus -0.07em}
   \sloppy \clubpenalty=4000 \widowpenalty=4000 \sfcode`\.=1000\relax}

\def\@cite#1#2{$[{{#1\if@tempswa , #2\fi}}]$}

\newcount\@tempcntc
\def\@citex[#1]#2{\if@filesw\immediate\write\@auxout{\string\citation{#2}}\fi
  \@tempcnta\z@\@tempcntb\m@ne\def\@citea{}\@cite{%
        \@ordonner{#2}%
        \@for\@citeb:=#2\do%
    {\@ifundefined{b@\@citeb}%
        {\@citeo\@tempcntb\m@ne\@citea%
                \def\@citea{,\penalty\@m\ }{\bf ?}\@warning%
                {Citation `\@citeb' on page \thepage \space undefined}}%
        {\setbox\z@\hbox{\global\@tempcntc0\csname b@\@citeb\endcsname\relax}
     \ifnum\@tempcntc=\z@ \@citeo\@tempcntb\m@ne%
       \@citea\def\@citea{,\penalty\@m}%
       \hbox{\csname b@\@citeb\endcsname}%
     \else%
      \advance\@tempcntb\@ne%
      \ifnum\@tempcntb=\@tempcntc%
      \else\advance\@tempcntb\m@ne\@citeo%
      \@tempcnta\@tempcntc\@tempcntb\@tempcntc\fi\fi}}\@citeo}{#1}}%

\def\@citeo{\ifnum\@tempcnta>\@tempcntb\else\@citea
  \def\@citea{,\penalty\@m}%
  \ifnum\@tempcnta=\@tempcntb\the\@tempcnta\else
   {\advance\@tempcnta\@ne\ifnum\@tempcnta=\@tempcntb \else
\def\@citea{-}\fi
    \advance\@tempcnta\m@ne\the\@tempcnta\@citea\the\@tempcntb}\fi\fi}

\def\@toto{}
\newif\if@ordre 
\newcount\c@current
\newcount\c@last

\def\@ordonner#1{\global\c@last\m@ne%
                \global\@ordretrue%
                \@for\@toto:=#1\do%
                        {\@ifundefined{b@\@toto}%
                        {}%
                        {\c@current\csname b@\@toto\endcsname\relax%
                        \ifnum\the\c@current<\the\c@last\relax%
                                {\global\@ordrefalse}\fi%
                        \global\c@last\the\c@current%
                        }%
                        }%
                \if@ordre{}\else{\typeout{}%
                        \typeout{Warning: the references are not %
                         in increasing order\on@line:}%
                        \@for\@toto:=#1\do%
                        {\@ifundefined{b@\@toto}%
                        {}%
                        \typeout{\@toto:\space \@nameuse{b@\@toto}}%
                        }\typeout{}}\fi%
                }%

\catcode`\@=12

\catcode`\@=11

\newcount\c@subequation

\def\eqnarray{
\def\@eqnnum{{\reset@font\rm%
(\theequation-{\alph{subequation}})}}
\global\c@subequation=1\relax
\stepcounter{equation}\let\@currentlabel\theequation
\global\@eqnswtrue\m@th
\global\@eqcnt\z@\tabskip\@centering\let\\\@eqncr
$$\halign to\displaywidth\bgroup\@eqnsel\hskip\@centering
  $\displaystyle\tabskip\z@{##}$&\global\@eqcnt\@ne
  \hskip 2\arraycolsep \hfil${##}$\hfil
  &\global\@eqcnt\tw@ \hskip 2\arraycolsep $\displaystyle\tabskip\z@{##}$\hfil
   \tabskip\@centering&\llap{##}\tabskip\z@\cr}

\def\@@eqncr{\let\@tempa\relax
    \ifcase\@eqcnt \def\@tempa{& & &}\or \def\@tempa{& &}%
      \else \def\@tempa{&}\fi
     \@tempa \if@eqnsw\@eqnnum\global\advance\c@subequation by 1\relax
                        \fi
     \global\@eqnswtrue\global\@eqcnt\z@\cr}

\def\endeqnarray{\@@eqncr\egroup
      \global\advance\c@equation\m@ne$$\global\@ignoretrue
        \stepcounter{equation}
        \def\@eqnnum{{\reset@font\rm (\theequation)}}}

\def\Eqnarray{
\def\@eqnnum{{\reset@font\rm (\theequation)}}
\global\c@subequation=1\relax
\stepcounter{equation}\let\@currentlabel\theequation
\global\@eqnswtrue\m@th
\global\@eqcnt\z@\tabskip\@centering\let\\\@eqncr
$$\halign to\displaywidth\bgroup\@eqnsel\hskip\@centering
  $\displaystyle\tabskip\z@{##}$&\global\@eqcnt\@ne
  \hskip 2\arraycolsep \hfil${##}$\hfil
  &\global\@eqcnt\tw@ \hskip 2\arraycolsep $\displaystyle\tabskip\z@{##}$\hfil
   \tabskip\@centering&\llap{##}\tabskip\z@\cr}

\def\endEqnarray{\@@eqncr\egroup
      \global\advance\c@equation\m@ne$$\global\@ignoretrue
        \stepcounter{equation}
        \def\@eqnnum{{\reset@font\rm (\theequation)}}}

\catcode`\@=12

\font\tenmsa=msam10
\font\sevenmsa=msam7
\font\fivemsa=msam5
\font\tenmsb=msbm10
\font\sevenmsb=msbm7
\font\fivemsb=msbm5
\newfam\msafam
\newfam\msbfam
\textfont\msafam=\tenmsa  \scriptfont\msafam=\sevenmsa
\scriptscriptfont\msafam=\fivemsa
\textfont\msbfam=\tenmsb  \scriptfont\msbfam=\sevenmsb
\scriptscriptfont\msbfam=\fivemsb

\global\mathchardef\lesssim "142E

\newcommand{\slL}{\raise.15ex\hbox{$/$}\kern-.53em\hbox{$L$}}
\newcommand{\slP}{\raise.15ex\hbox{$/$}\kern-.53em\hbox{$P$}}
\newcommand{\slR}{\raise.15ex\hbox{$/$}\kern-.53em\hbox{$R$}}
\newcommand{\slQ}{\raise.15ex\hbox{$/$}\kern-.53em\hbox{$Q$}}
\newcommand{\slK}{\raise.15ex\hbox{$/$}\kern-.53em\hbox{$K$}}
\newcommand{\slSigma}{\raise.15ex\hbox{$/$}\kern-.53em\hbox{$\Sigma$}}
\newcommand{\slcalP}{\raise.15ex\hbox{$/$}\kern-.63em\hbox{$\cal P$}}

\newcommand{\be}{\begin{equation}}
\newcommand{\ee}{\end{equation}}     
\newcommand{\bea}{\begin{eqnarray}}
\newcommand{\ena}{\end{eqnarray}}

\def\build#1\over#2{\mathrel{\mathop{\kern 0pt#1}\limits_{#2}}}
\def\oldstyle#1{$\mit #1$}

\font\tenimbf=cmmib10 at 12pt
\font\sevenimbf=cmmib10 at 7pt
\font\fiveimbf=cmmib10 at 5pt
\newfam\imbf
\textfont\imbf=\tenimbf
\scriptfont\imbf=\sevenimbf
\scriptscriptfont\imbf=\fiveimbf
\def\imb{\fam\imbf\tenimbf}

\begin{document}
\begin{titlepage}
\title{\bf{Bremsstrahlung and photon production\\
 in thermal QCD}}
\author{
P.~Aurenche$^{(1)}$, F.~Gelis$^{(1)}$, 
R.~Kobes$^{(2)}$, H.~Zaraket$^{(1)}$}
\maketitle

\begin{center}
\begin{enumerate}
\item Laboratoire de Physique Th\'eorique LAPTH,\\
BP110, F-74941, Annecy le Vieux Cedex, France
\item Physics Department and Winnipeg Institute
for Theoretical Physics,\\
University of Winnipeg,
Winnipeg, Manitoba R3B 2E9, Canada
\end{enumerate}
\end{center}

\bibliographystyle{unsrt}
\vskip 1cm

\begin{abstract}
In this paper, we extend the study of bremsstrahlung photon production in a
quark-gluon plasma to the cases of soft static photons 
$({\imb q}={\imb 0})$ and hard real photons. 
The general framework of this study is the effective perturbative
expansion based on the resummation of hard thermal loops. Despite the
fact that bremsstrahlung only comes at two loops, we find that in both
cases it generates contributions of the same order of magnitude as
those already calculated by several other groups at one loop.
Furthermore, a new process contained in the two-loop diagrams dominate
the emission of a very hard real photon. In all cases, the rate of real
or virtual photon production in the plasma is appreciably increased compared to
the one-loop predictions.
\end{abstract}
\vskip 4mm
\centerline{\hfill LAPTH--678/98,  WIN--98-05, hep-ph/9804224\hglue 2cm}
\vfill
\thispagestyle{empty}
\end{titlepage}

\section{Introduction}
We consider the production of a real photon or of a lepton pair in a
quark-gluon plasma. The plasma is assumed to be in equilibrium at
temperature $T$. The theoretical framework used in the calculation is
that of thermal field theory improved by the hard loop resummation
\cite{Pisar2,Pisar6,BraatP1,BraatP2,FrenkT1,FrenkT2} of
Braaten and Pisarski: in this approach one distinguishes hard momenta,
of order $T$, from soft momenta, of order $g T$, where $g$ is the
Quantum Chromodynamics (QCD) coupling constant assumed to be small ($g
\ll 1$). After resummation of hard thermal loops, one is led to an
effective field theory from which observables can be evaluated
perturbatively.

The production rates of real or virtual photons have already been
evaluated, at the one loop level, in the effective theory
\cite{BraatPY1,BaierPS1,AurenBP1,BaierNNR1,KapusLS1}. Concerning soft
virtual photons, it was found that the rate of production is
considerably modified and enhanced compared to the result of the bare
theory. Besides the usual quark-antiquark annihilation process, there
appear many production mechanisms, in particular processes where the
photon is radiated off a (hard) quark in a scattering process where the
quark is backward scattered in the  plasma via soft quark exchange. This
is to be contrasted to the result obtained in a semi-classical
approximation \cite{CleymGR2,CleymGR3,GolovR1,BaierDMPS1,BaierDMPS2}
where the photon is radiated off fast quarks in scattering processes
mediated by a gluon exchange: we call such processes bremsstrahlung
emission of a photon.
In this study we reconcile the two approaches and show
that the bremsstrahlung processes favored by the semi-classical
approximation appear at the two-loop level in the effective theory and
that, in fact, they contribute at the same order in $g T$ as the
processes in the one-loop effective theory.  Such a result should not be
a surprise. 

Consider the case of a soft virtual photon. The rate of production is
related to the imaginary part of the vacuum polarization diagram
\cite{Weldo3,GaleK1}. In the one-loop approximation of the effective
field theory, it involves only effective fermion propagators and
effective vertices \cite{BraatPY1}. A dominant contribution to the rate
arises when the internal fermion momentum is soft and, therefore, all
effective propagators and effective vertices have the same order of
magnitude as their bare counterparts. A close examination of the final
result shows, however, that it has a logarithmic sensitivity to scales
of order $T$ (see section \ref{BPYcomp}):  this means that such a
diagram also receives a dominant contribution from hard fermion momenta.
When the momentum becomes large, the hard thermal loop (HTL) corrections
to propagators and vertices are suppressed by, at least, a factor $g$
with respect to their bare counterparts. This suppression factor $g$ can
easily be compensated by the larger phase space available to a hard
fermion (${\cal O}(T)$) compared to soft fermion (${\cal O}(g T)$),
thereby leading to a contribution of the same order of magnitude from
the soft region and the hard region of phase-space. Now, when an observable is
sensitive to the thermal corrections of hard vertices and propagators, it is
obvious that all such corrections should be taken into account for the
calculation to be complete. Some of these thermal corrections are naturally
included in the lowest order of the effective theory via the resummation of hard
thermal loops. But, even if the HTL approximation is correct for soft external
particles, it does not account for all thermal corrections to hard vertices and
hard propagators. For instance, neglecting the external momenta as one does in
the calculation of the hard thermal loops is no longer justified when these
momenta are not soft. Besides, equally important may be the contribution arising
from soft gluons in the loop giving the HTL when the external momenta are hard,
due to the Bose enhancement of the soft gluon term. Within the effective theory,
both types of additional thermal corrections to a hard propagator or vertex are
taken into account by considering a one-loop correction to this propagator or
vertex.

In the calculation of the virtual photon production rate in the
effective theory, soft gluon exchange appears in two-loop diagrams. It
will be seen that the bremsstrahlung production mechanism is precisely
given by these diagrams when the exchanged gluon is space-like. The
evaluation  of these diagrams is discussed below. These contributions
are clearly not included in the effective one-loop diagram. This is
obvious when looking at the corresponding physical processes and it
manifests itself in the result by the calculated rate being proportional
to the square of the thermal gluon mass  $m_{\rm g}$, in contrast to the
one-loop result where only the thermal quark mass appears
\cite{BraatPY1}. 
Another important contribution of the
two-loop diagrams is associated with time-like gluon exchanges: physically
this represents QCD Compton scattering and quark-antiquark annihilation to
produce a gluon and a photon. To evaluate this properly requires care
since the two-loop diagrams with a hard time-like gluon exchanged are
already part of the one-loop diagram with effective propagators and
vertices. Taking into account the contribution of the appropriate
counterterms in the effective lagrangian will prevent double-counting
and allow the correct evaluation of the soft, time-like, gluon
contribution.

The case of soft real photon production follows essentially the same
pattern, except for the crucial fact that, the external line in the
vacuum polarization diagram being massless, collinear divergences appear
when evaluating the two-loop diagrams: the quasi-overlap of two such
divergences, associated with the fermion propagators, leads to an
enhancement factor of ${\cal O} (1/g^2)$ \cite{AurenGKP2,AurenGKP1}. 
The paradoxical result then follows that the one-loop contribution is
relatively suppressed by a factor $g^2$ compared to the two-loop one!
The latter is entirely dominated by the bremsstrahlung process since the
kinematical constraints require the gluon to be space like for the
enhancement factor to occur. The Compton and annihilation mechanisms are
sub-dominant.

The case of a hard real photon, of momentum of ${\cal O} (T)$, shares
features with both cases above. The one-loop approximation has a
logarithmic sensitivity to the hard fermion momentum in the loop
\cite{BaierNNR1,KapusLS1}. The
two-loop bremsstrahlung has a $1/g^2$ collinear enhancement, as in the
real soft photon case, which however is compensated by a factor 
$m_{\rm g}^2/q^2 \sim g^2$, where $q$ is the photon momentum, 
leaving the bremsstrahlung
contribution at the same order in $g T$ as the one-loop contribution.

In the following we are concerned mainly with the bremsstrahlung part of
the two-loop diagrams, and leave the discussion of the Compton and
annihilation processes and their interplay with the counterterms of the
effective theory to future work. We do not discuss the production of
soft real photons since this has already been studied in detail in 
\cite{AurenGKP2}.

In the next section we derive the general expression for the (real or
virtual) photon production rate at the two loop level. Then we consider
the case of soft virtual photons produced at rest in the plasma  and
derive the leading behavior analytically. We compare to the one-loop
results and show that the bremsstrahlung contribution is numerically
dominant although both contributions are technically of the same order
in $g$. The semi-classical approach is then discussed and it is shown
that even though the approximations inherent in the semi-classical
approach are not really justified in the case of soft photon production
in a quark-gluon plasma,  it leads to a result quite comparable to that
obtained in thermal field theory. Turning to the case of hard real
photons, it is shown that the bremsstrahlung mechanism is of the same
order as the already calculated one-loop result. Carrying out a more
detailed comparison with the latter approach, it is found that the
bremsstrahlung process dominates over the one-loop result 
for photon momentum of ${\cal O} (T)$
but is relatively suppressed by a logarithmic factor for hard enough photons. 
For very hard energies, we find that the photon production is in fact
dominated by a new process consisting of a $q\bar{q}$ annihilation where
the quark or antiquark undergoes a scattering in the medium.  We summarize all the
thermal field theory results concerning real and virtual photon
production in a concluding section.

The role of counterterms in the application of the effective theory up
to two-loops is discussed in a first appendix where the problem is also
illustrated in a simple example. In a second appendix, the importance of
phase space factors in thermal calculations is emphasized and the
difference with the zero temperature phase space is made clear.

\section{Bremsstrahlung in thermal field theory}
\label{sec:bremss}
\subsection{Topologies involving bremsstrahlung}
Let us first recall the relationship between the photon production rate
and the imaginary part of the retarded 
polarization tensor of the photon, as
given by thermal field theory (we follow  the notations of \cite{EijckKW1}). 
For real photons, this relation gives the
number of photon emitted per unit time and per unit volume of the plasma
as \cite{Weldo3,GaleK1}:
\begin{equation}
  {{dN}\over{dtd{\imb x}}}=-
  {{d{\imb q}}\over{(2\pi)^3 2q_o}}\;2n_{_{B}}(q_o)\,
  {\rm Im}\,\Pi^{^{RA}}{}_\mu{}^\mu(q_o,{\imb q})\; ,
  \label{realphot}
\end{equation}
whereas for the production of a photon of invariant mass $Q^2$ decaying
into a lepton pair we have:
\begin{equation}
  {{dN}\over{dtd{\imb x}}}=-
  {{dq_od{\imb q}}\over{12\pi^4}}\;
  {\alpha\over{Q^2}}\,n_{_{B}}(q_o)\,
  {\rm Im}\,\Pi^{^{RA}}{}_\mu{}^\mu(q_o,{\imb q})\; .
  \label{virtualphot}
\end{equation}
Basically, the above two formulae differ only by the allowed phase
space for the photon, by an extra QED coupling constant when the photon
decays into a lepton pair, and by the propagator of a heavy photon. It
is worth recalling that these relations are valid only at first order
in the QED coupling $\alpha$ since they do not take into account the
possible re-interactions of the photon on its way out of the plasma
nor the simultaneous emission of more than one photon.
Nevertheless, they are true to all orders in the strong coupling
constant $\alpha_{_{S}}$. This should not be a serious limitation from
a practical point of view since $\alpha\ll\alpha_{_{S}}$. 

\begin{figure}[htbp]
  \centerline{
    \resizebox*{!}{2.5cm}{\includegraphics{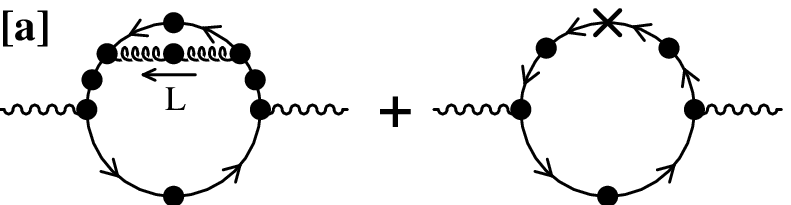}}
    }
  \vskip 2mm
  \centerline{
    \resizebox*{!}{2.5cm}{\includegraphics{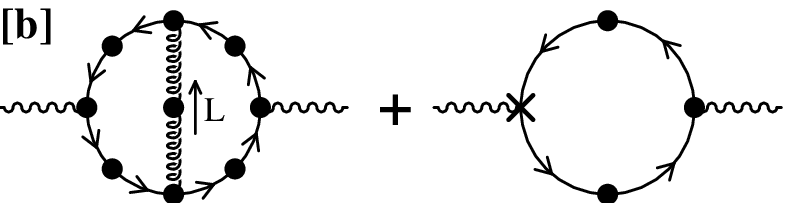}}
    }
  \caption{\footnotesize{Two-loop contributions involving 
      bremsstrahlung processes. A black dot denotes an effective
      propagator or vertex. Crosses are HTL counterterms.}}
  \label{fig:2loop}
\end{figure}

Let us now examine in which topologies the bremsstrahlung can appear. It
is worth recalling at this point that the retarded imaginary part in
Eqs.~(\ref{realphot}) and (\ref{virtualphot}) 
can be expressed as a sum over possible cuts
through the diagram \cite{KobesS1,KobesS2,Gelis3}. 
Therefore, we need to look for
diagrams that will give bremsstrahlung processes once cut. A simple
inspection of the processes involved in one-loop contributions (see
\cite{BraatPY1,AurenGKP2} for instance) shows that bremss\-tra\-hlung 
does not appear at this order. To see
bremsstrahlung processes, one should consider the two-loop
contributions of Fig.~\ref{fig:2loop}. 
The diagrams have been obtained via a strict application of the
Feynman rules of the effective theory \cite{BraatP1,FrenkT1}, giving
{\it a priori} effective vertices and propagators and
diagrams with counterterms\footnote{These
counterterms are nothing but the HTL contribution to the two or
three-point function, with the opposite sign. Formally, they are
necessary
because one wants the effective theory to be just a reordering of the
bare perturbative expansion, with the same overall Lagrangian.} 
in order to avoid
any double counting of thermal corrections already
included at the one-loop
level via the resummation of hard thermal loops, as outlined in appendix
\ref{app:counterterms}. 
To make the connection with previous works
\cite{AurenGKP2,AurenGKP1} easier, we mention that looking at two
loop diagrams in the effective theory is just a more rigorous way of
doing what we might call ``calculating one-loop diagrams beyond the HTL
approximation''. Our present formulation is indeed more rigorous 
since it takes
care of the counterterms, and also more positive since it does not
assume {\it a priori} that one needs to go beyond the effective 
theory.
Among all the possible
cuts through the diagrams, those that correspond to bremsstrahlung
necessarily cut the gluon propagator. Moreover, if $L$ is the gluon
4-momentum, only the Landau damping part $(L^2<0)$ gives bremsstrahlung,
the $L^2>0$ part rather giving Compton effect or $q\bar{q}$-like
annihilations \cite{WorkIP1}.
There is another reason why the region $L^2>0$ deserves a separate
treatment:
in the $L^2<0$ kinematical
domain, it is obvious that we cannot have contributions coming from the
HTL counterterms since these counterterms involve only bare gluon
propagators that don't have any imaginary part in the space-like region.
On the contrary, in the
$L^2>0$ region, one should pay special attention to the counterterm
diagrams in order to avoid any double-counting. Indeed, when the gluon
becomes hard, we have a hard loop that may reproduce what is already
included in the one-loop diagram via the effective vertices and
propagators.
From now on, we limit ourselves to the region where
$L^2<0$ and only to the true two-loop diagrams, leaving 
the $L^2>0$ region and the discussion of counterterms to future work.

Moreover, since our main
focus is on bremsstrahlung, we must retain from the cut quark
propagators only the pole part and reject the Landau damping part, which would
correspond to a different physical process. For the same
reason, the cut should avoid going through an effective
vertex\footnote{These extra requirements are not a claim that other
configurations of the cut cannot give important contributions as well,
but are dictated by our choice of looking only at bremsstrahlung.}.

\subsection{General expression of two-loop contributions}
\begin{figure}[htbp]
  \centerline{
    \resizebox*{!}{2.5cm}{\includegraphics{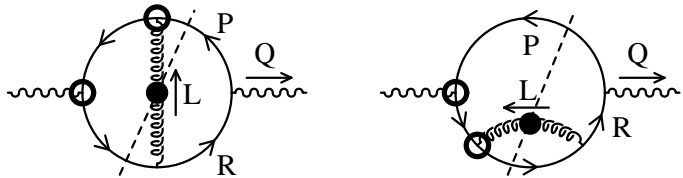}}}
  \caption{\footnotesize{Simplified two-loop contributions involving 
      bremsstrahlung processes. The circled vertices correspond to the
      framework of the cutting rules.}}
  \label{fig:2loopsimple}
\end{figure}

In order to obtain a bremsstrahlung contribution of the
same order of magnitude as the already calculated one-loop contributions, we
need a hard phase space for the quark circulating in the loop, as
explained in appendix \ref{app:phasespace}. We will see later that we
have such contributions in the diagrams depicted in
Fig.~\ref{fig:2loop}. Therefore, at leading order, we can 
use ``bare''\footnote{As in
\cite{AurenGKP2,AurenGKP1,FlechR1}, we may have to keep an asymptotic
thermal mass even in the hard region to regularize collinear divergences
encountered when the external photon is on--shell. But contrary to
\cite{FlechR1}, we don't need to add this mass by hand, since it is naturally
contained in the hard limit of effective quark propagators which are our starting
point in Fig.~\ref{fig:2loop}.} vertices and propagators everywhere
except for the gluon propagator
since the gluon can be soft. The diagrams we have to consider are
therefore the simplified versions of the previous ones represented on
Fig.~\ref{fig:2loopsimple}.
In the same figure, we have depicted the relevant cuts as well as the
arrangement of circlings that enables one to calculate the corresponding
contribution in the framework of the thermal cutting rules. We have checked 
that the two cuts represented form a gauge independent set of
terms, to which one should add the symmetric cut for the vertex diagram
and a third diagram with the self-energy correction on the upper quark
line. Since these two other terms give the same contribution as the
previous two, we simply take them into account by multiplying the final
result by an overall factor of 2.

A straightforward application of the cutting rules valid for the ``R/A''
formalism, with the notations of \cite{Gelis3}, 
gives for the vertex correction:
\begin{Eqnarray}&&\!\!\!
  {\rm Im}\,\Pi^{^{RA}}
  {}_\mu{}^\mu(q_o,{\imb q})_{|_{\rm vertex}}\!\!\!=-
  {\rm Im}\,\Pi^{^{AR}}
  {}_\mu{}^\mu(q_o,{\imb q})_{|_{\rm vertex}}=
  {{NC_{_{F}}}\over 2}
  \int{{d^4P}\over{(2\pi)^4}}\int{{d^4L}\over{(2\pi)^4}}
  \nonumber\\
  &&\times\;
  e^{^{\underline{ARR}}}(Q,P+L,-R-L) g^{^{\underline{ARR}}}(-P-L,P,L)
  \nonumber\\
  &&\times\;
  g^{^{{AAR}}}(R+L,-L,-R) 
  e^{^{{AAR}}}(R,-P,-Q)
  \nonumber\\
  &&\times\;
  {\rm Tr}\,\left[\gamma^\mu{\cal S}^{^{\underline{AR}}}(P+L)\gamma^\rho
    {\cal S}^{^{A\underline{R}}}(P)\gamma_\mu\right.
  \left.{\cal S}^{^{RA}}(R)\gamma^\sigma
    {\cal S}^{^{\underline{R}A}}(R+L)\right]
  D^{^{A\underline{R}}}_{\rho\sigma}(L)
  \nonumber\\
  &&=-{{NC_{_{F}}}\over 2}e^2g^2
  \int{{d^4P}\over{(2\pi)^4}}\!\!\!\int{{d^4L}\over{(2\pi)^4}}
  \left[\Delta^{^{R}}_{_{T,L}}(L)\!-\!\Delta^{^{A}}_{_{T,L}}(L)\right]
  \nonumber\\
  &&\times\;
  \left[S^{^{R}}(P)-S^{^{A}}(P)\right]
  \left[S^{^{R}}(R+L)-S^{^{A}}(R+L)\right]
  \nonumber\\
  &&\times\;
  \left(n_{_{F}}(r_o)-n_{_{F}}(p_o)\right)
  \left(n_{_{B}}(l_o)+n_{_{F}}(r_o+l_o)\right)
  \nonumber\\
  &&\times\;
  S(R)S(P+L)P_{\rho\sigma}^{^{T,L}}(L)\;
  {\rm Trace}^{\rho\sigma}{}_{|_{\rm vertex}}\; ,
  \label{cutvertex}
\end{Eqnarray}
where, following \cite{AurenGKP2}, we denote the fermion propagator:
\begin{eqnarray}
  &&{\cal S}^{^{R,A}}(P)\equiv{\overline{\slP}} S^{^{R,A}}(P){\ \rm with\ }
  {\overline{P}}\equiv(p_o,\sqrt{p^2+M^2_\infty}\,\hat{\imb p})\\
  &&S(P)^{^{R,A}}
  \equiv{i\over{{\overline{P}}^2\pm ip_o\varepsilon}}
  ={i\over{P^2-M^2_{\infty}\pm ip_o\varepsilon}}\; ,
  \label{notations:quark}
  \end{eqnarray}
 and the effective gluon propagator in a linear covariant gauge:
 \begin{eqnarray}
  &&-D_{\rho\sigma}^{^{R,A}}(L)\equiv
  P^{^{T}}_{\rho\sigma}(L)\Delta^{^{R,A}}_{_{T}}(L)
  +P^{^{L}}_{\rho\sigma}(L)\Delta^{^{R,A}}_{_{L}}(L)
  +\xi L_\rho L_\sigma/L^2\\
  &&\Delta^{^{R,A}}_{_{T,L}}(L)\equiv
  \left.{i\over{L^2-\Pi_{_{T,L}}(L)}}\right|_{_{R,A}}
  {\ \rm and\ \ } 
  \rho_{_{T,L}}(L)\equiv {\rm Disc}\Delta^{^{R}}_{_{T,L}}(L)
  \\
  &&\Pi_{_{T}}(L)\equiv3m_{\rm g}^2\left[
    {{x^2}\over 2}+{{x(1-x^2)}\over{4}}\ln\left(
      {{x+1}\over{x-1}}\right)\right]\\
  &&\Pi_{_{L}}(L)\equiv3m_{\rm g}^2(1-x^2)\left[
    1-{{x}\over{2}}\ln\left(
      {{x+1}\over{x-1}}\right)\right]\; ,
  \label{notations:gluon}
\end{eqnarray}
\noindent
with $P^{^{T,L}}_{\rho\sigma}$ the usual transverse and longitudinal
projectors in linear covariant gauges 
\cite{Pisar6,Weldo1,LandsW1,Klimo1,Klimo2},
$M^2_{\infty}\equiv g^2 C_{_{F}} T^2/4$ \cite{FlechR1} the asymptotic
thermal mass of the quark, and $m_{\rm g}^2
\equiv g^2T^2[N+N_{_{F}}/2]/9$ the soft gluon thermal mass. 
In this formula, $e$ is the electric
charge of the quark and therefore depends on its flavor. Likewise, we
obtain for the second diagram:
\begin{Eqnarray}
  &&
  \!\!\!{\rm Im}\,\Pi^{^{RA}}
  {}_\mu{}^\mu(q_o,{\imb q})_{|_{\rm self}}=-
  {\rm Im}\,\Pi^{^{AR}}
  {}_\mu{}^\mu(q_o,{\imb q})_{|_{\rm self}}=
  {{NC_{_{F}}}\over 2}
  \int{{d^4P}\over{(2\pi)^4}}\int{{d^4L}\over{(2\pi)^4}}
  \nonumber\\
  &&\times\;
  e^{^{\underline{ARR}}}(Q,P,-R)
  g^{^{\underline{ARR}}}(R,L,-R-L)
  \nonumber\\
  &&\times\;
  g^{^{{AAR}}}(R+L,-L,-R) e^{^{{AAR}}}(R,-P,-Q)
  \nonumber\\
  &&\times\;
  {\rm Tr}\,\left[\gamma^\mu{\cal S}^{^{A\underline{R}}}(P)\gamma_\mu
    {\cal S}^{^{{RA}}}(R)\gamma^\rho\right.
  \left.{\cal S}^{^{\underline{R}A}}(R+L)\gamma^\sigma
    {\cal S}^{^{\underline{RA}}}(R)\right]
  D^{^{A\underline{R}}}_{\rho\sigma}(L)
  \nonumber\\
  &&=-{{NC_{_{F}}}\over 2}e^2g^2
  \int{{d^4P}\over{(2\pi)^4}}\int{{d^4L}\over{(2\pi)^4}}
  \left[\Delta^{^{R}}_{_{T,L}}(L)-\Delta^{^{A}}_{_{T,L}}(L)\right]
  \nonumber\\
  &&\times\;
  \left[S^{^{R}}(P)-S^{^{A}}(P)\right]
  \left[S^{^{R}}(R+L)-S^{^{A}}(R+L)\right]P_{\rho\sigma}^{^{T,L}}(L)\;
  {\rm Trace}^{\rho\sigma}{}_{|_{\rm self}}
  \nonumber\\
  &&\times\;
  \left(S(R)\right)^2
  \left(n_{_{F}}(r_o)-n_{_{F}}(p_o)\right)
  \left(n_{_{B}}(l_o)+n_{_{F}}(r_o+l_o)\right)\; .
  \label{cutself}
\end{Eqnarray}
\noindent
In the previous formulae, a factor $S(R)$ without any $R$ or $A$ 
superscript simply
denotes the principal part of the propagator. In other words, for these
factors, the $R$ or $A$ prescription is irrelevant because the corresponding
delta function is incompatible with the other delta functions present
and therefore vanishes.

We may notice the similarity between Eqs.~(\ref{cutvertex}) and (\ref{cutself}).
In particular, the same combination of statistical weights appear in both
formulae, while the expressions in the square brackets simply express the cuts
on internal lines. Moreover, when plugged into
Eq.~(\ref{realphot}) in order to obtain the production rate, the sum of
Eqs.~(\ref{cutvertex}) and (\ref{cutself}) gives the more intuitive
Eq.~(\ref{eq:lastform}) in appendix \ref{app:phasespace} (to which one should
add similar terms to take into account all the processes included in our
diagrams). Therefore, the ``R/A'' formalism appears just as an efficient method
to reorder the factors of the integrand in order to make it more compact and
more convenient for the subsequent integrations. The drawback of this formalism
is that it generates less intuitive expressions.

\subsection{Common part of the calculation}
The calculation of the Dirac's traces is of course common to both cases.
We obtain for the self-energy insertion:
\begin{Eqnarray}
  &&{\rm Trace}{}^{\rho\sigma}{}_{|_{\rm self}}
  \approx
  -4\left[ 4{\overline{R}}{}^2Q{}^\rho {\overline{R}}{}^\sigma 
    -4Q{}^2{\overline{R}}{}^\rho {\overline{R}}{}^\sigma\right.
  \nonumber\\
  &&\qquad\qquad\left.-g{}^{\rho\sigma}\left(
      {\overline{R}}{}^2({\overline{R}}{}^2-Q{}^2)
      +2{\overline{R}}{}^2Q\cdot L-2Q{}^2
      {\overline{R}}\cdot L\right)\right]\; ,
\end{Eqnarray}
and for the vertex correction:
\begin{Eqnarray}
  &&{\rm Trace}{}^{\rho\sigma}{}_{|_{\rm vertex}}
  \approx
  -4\left[2{\overline{R}}{}^2{\overline{P}}{}^\rho Q{}^\sigma
    -2({\overline{P+L}}){}^2{\overline{R}}{}^\rho Q{}^\sigma
  \right.
  \nonumber\\
  &&
  \qquad
  +2L{}^2({\overline{R}}{}^\rho {\overline{R}}{}^\sigma
  +{\overline{P}}{}^\rho {\overline{P}}{}^\sigma)
  -4Q{}^2{\overline{R}}{}^\rho {\overline{P}}{}^\sigma
  \nonumber\\
  &&\qquad+g{}^{\rho\sigma}\left(
    \left.-L{}^2({\overline{R}}{}^2+({\overline{P+L}}){}^2-Q{}^2-L{}^2)
    \right)\right]\; .
\end{Eqnarray}
It is worth recalling that these expressions are obtained by
anticipating the use
of the relation
\begin{equation}
  L^\rho P^{^{T,L}}_{\rho\sigma}(L)=0
\end{equation}
in order to drop any $L^\rho$ or $L^\sigma$ in the expression of the
Dirac's traces. Since this identity is not true for the gauge dependent
part of the gluon propagator, one should not use these expressions of
the traces to check the independence of the rate with respect to the
gauge parameter $\xi$. Moreover, we discarded terms that will be killed
later by the delta functions such as the one contained in
$S^{^{R}}(P)-S^{^{A}}(P)=2\pi\epsilon(p_o)\delta(P^2-M^2_\infty)$. 
Since a 4-vector like 
${\overline{P}}=(p_o,\sqrt{p^2+M^2_\infty}\,\hat{\imb p})$ is not a linear
function of the momentum $P$, we used some approximations to simplify
the calculations, the effect of which 
is to neglect only terms that are
always subdominant\footnote{For instance:
\begin{equation}
\overline{R}-\overline{P}=Q(1+{\cal O}\left({{M^2_\infty}\over{p(p+q)}}
\right))\; .
\end{equation}}.

We also notice that the statistical weights and delta functions present in
Eqs.~(\ref{cutvertex}) and (\ref{cutself}) are invariant under the change of
variables $P\to -R-L$, $L\to L$. Therefore, in the remaining factors of the
integrand, we are allowed to drop the parts which are antisymmetric under this
transformation. Collecting contributions from the two topologies, this
symmetrization gives:
\begin{Eqnarray}
  &&\left[{{{\rm Trace}{}^{\rho\sigma}{}_{|_{\rm vertex}}}
    \over{\overline{R}{}^2(\overline{P+L}){}^2}}
  +{{{\rm Trace}{}^{\rho\sigma}{}_{|_{\rm self}}}
    \over{\overline{R}{}^2\overline{R}{}^2}}\right]_{\rm sym}\approx-4\left[
    2L{}^2{{\overline{R}{}^\rho\overline{R}{}^\sigma+
      \overline{P}{}^\rho\overline{P}{}^\sigma}
    \over{\overline{R}{}^2(\overline{P+L}){}^2}}
    \right.\nonumber\\
    &&\qquad
  -2{Q{}^2}\left({{\overline{R}{}^\rho}\over{\overline{R}{}^2}}
  +{{\overline{P}{}^\rho}\over{(\overline{P+L}){}^2}}\right)
  \left({{\overline{R}{}^\sigma}
      \over{\overline{R}{}^2}}
    +{{\overline{P}{}^\sigma}
      \over{(\overline{P+L}){}^2}}\right)
  \nonumber\\
  &&\qquad -g{}^{\rho\sigma}\left(
    1
    -{{L{}^2(L{}^2+Q{}^2)}\over{\overline{R}{}^2(\overline{P+L}){}^2}}
    +({{L^2+Q\cdot L}})\left({1\over{\overline{R}{}^2}}
  +{1\over{(\overline{P+L}){}^2}}\right)
  \right.
  \nonumber\\
  &&\qquad \left.\left.
      +{{Q{}^2L{}^2}\over{2}}\left({{1}\over{(\overline{R}{}^2)^2}}+
      {{1}\over{((\overline{P+L}){}^2)^2}}\right)
    \right)\right]\; ,
  \label{eq:tracetotal}
\end{Eqnarray}
which will be our starting point in the following sections.

\section{Contribution to soft static photons}
\subsection{Kinematics}
When the 3-momentum ${\imb q}$ of the emitted photon is zero, a lot of
simplifications occur. First of all, since there is one 3-vector less in
the problem, we need only one angular variable which simplifies
considerably the angular integrations. Moreover, as shown in
\cite{AurenGKP2}, the non vanishing invariant mass of the emitted photon
regulates all the potential collinear divergences when $Q^2/q_o^2\sim
1$. Therefore, one can simply forget about the quark
asymptotic thermal mass since the purpose of such a mass is precisely to
regularize collinear singularities. This means that we can everywhere
identify $P$ and $\overline{P}$ at this level of approximation, since
furthermore $P$ is hard.

From the identity
$S^{^{R}}(P)-S^{^{A}}(P)=2\pi\epsilon(p_o)\delta(P^2)$, we extract the
values $p_o=\pm p$ and $r_o=q_o\pm p$. For the second cut quark
propagator, we  have the identity 
$S^{^{R}}(R+L)-S^{^{A}}(R+L)=2\pi\epsilon(r_o+l_o)\delta((R+L)^2)$ from
which we extract the cosine of the angle $\theta^\prime$
between ${\imb p}$ and ${\imb l}$:
\begin{equation}
  \cos\theta^\prime={{(r_o+l_o)^2-p^2-l^2}\over{2pl}}\; .
  \label{eq:cosine}
\end{equation}
Of course, we must require that this value be in $[-1,+1]$, which will
reduce the available phase space.
\begin{figure}[t]
  \centerline{\resizebox*{!}{5cm}{\includegraphics{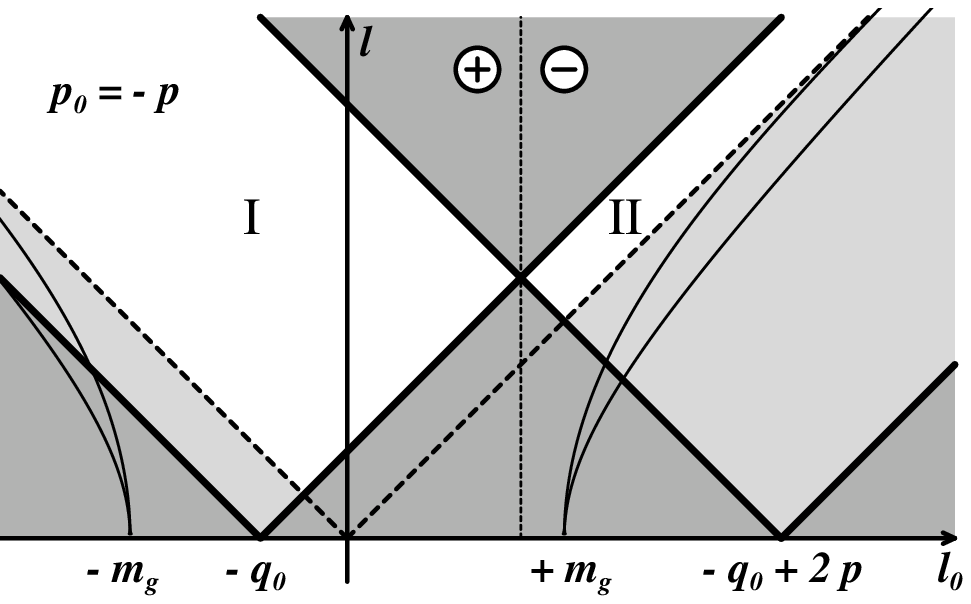}}}
  \centerline{\resizebox*{!}{5cm}{\includegraphics{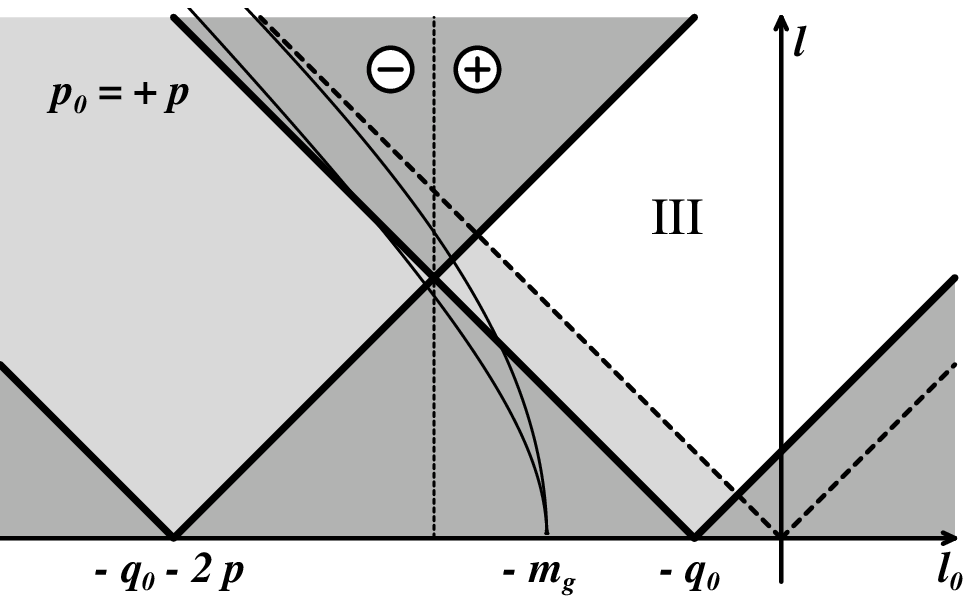}}}
  \caption{\footnotesize{Allowed domains in the $(l_o,l)$ plane 
      for $p_o=\pm p$. 
      The area shaded in dark
      gray is excluded by the delta functions. The region shaded in light
      gray is above the light--cone (dotted lines). 
      The solid curves are the transverse and
      longitudinal dispersion curves of the thermalized gluon. The vertical 
      dotted
      line is the separation between $\epsilon(p_o)\epsilon(r_o+l_o)=+1$ and 
      $\epsilon(p_o)\epsilon(r_o+l_o)=-1$. The value of the thermal mass has
      been exaggerated in order to make the figure more readable.}}
  \label{fig:kine}
\end{figure}
This requirement leads to the following two inequalities:
\begin{eqnarray}
  &&(l_o-l+p_o+q_o-p)(l_o+l+p_o+q_o+p)\le 0 \\
  &&(l_o-l+p_o+q_o+p)(l_o+l+p_o+q_o-p)\ge 0\; ,
\end{eqnarray}
which lead to a phase space reduction that can be seen in
Fig.~\ref{fig:kine}, where the region excluded by the requirement
$\cos\theta^\prime\in[-1,+1]$ has been shaded in dark gray. Other
regions are excluded also by our choice of looking only at
bremsstrahlung, {\it i.e.} excluding areas where $L^2>0$ 
\cite{WorkIP1}. Finally, the
only regions we have to consider are the unshaded ones.

Having taken these constraints into account, the independent variables
we are left with are for instance (the choice is not unique) $r=p$,
$l_o$ and $l$, everything else being a function of these three.
In particular, the denominators appearing in the calculations are
\begin{eqnarray}
  &&R^2=q_o(q_o\pm 2p)\\
  &&(P+L)^2=-q_o(q_o+2l_o\pm2p)\; ,
  \label{eq:denos}
\end{eqnarray}
the $+$ sign corresponding to $p_o=+p$ and the $-$ sign to $p_o=-p$.

\begin{figure}[htbp]
  \centerline{
    \resizebox*{!}{4cm}{\includegraphics{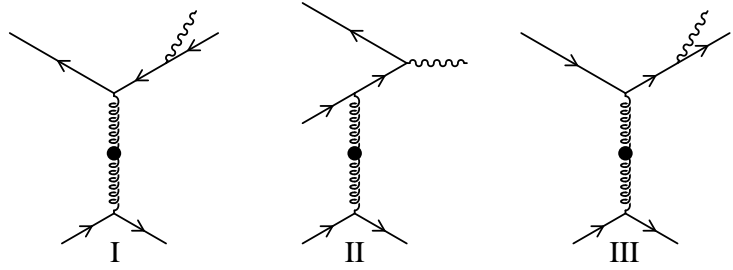}}
    }
  \caption{\footnotesize{Physical processes included in the diagrams of
      Fig.~\ref{fig:2loopsimple}, in the region $L^2<0$.
      Region I: $p_o<0$, $r_o+l_o<0$: bremsstrahlung with an
      antiquark. Region II: $p_o<0$, $r_o+l_o>0$: $q\bar{q}$ annihilation
      with scattering. Region III: 
      $p_o>0$, $r_o+l_o>0$: bremsstrahlung with a quark. The particle on which
      the quark is scattered can also be a gluon.}}
  \label{fig:processes}
\end{figure}

It is worth examining more closely 
to which physical processes the regions I, II and III correspond.
This is done just by looking at the
signs of $p_o$ and $r_o+l_o$. Examples are shown in
Fig.~\ref{fig:processes}. 
Besides the bremsstrahlung present in regions
I and III, we see a new process in region II. This process corresponds
to an annihilation of a quark-antiquark pair, one of the particles
having previously undergone a scattering in the medium. Since the scale
for the quark momentum is given by the temperature $(p={\cal O}(T))$,
and since we are looking here at soft photons, it is obvious from
Fig.~\ref{fig:kine} that processes I and III have support of order
$T^2$ in the $(l_o,l)$ plane while process II has only support of
order $q_o T$, the integrand being the same. Therefore, we expect and we have
checked that
process II is suppressed by a factor of order $q_o/T\ll 1$ compared
to bremsstrahlung. As a consequence, bremsstrahlung appears to be the
dominant contribution as far as the $L^2<0$ domain is concerned. In the
remaining part of this section, we limit our study to the regions I and III. We
can obtain a further reduction of the phase space by noticing that
regions I and III give the same contribution since they are equivalent
by a change of variables (indeed, after the symmetrization in
Eq.~(\ref{eq:tracetotal}), 
the integrands  are invariant under the change
of variables $P\to -R-L$ and $L\to L$). Physically, this means that photons are
produced equally by quarks and by antiquarks.
Therefore, we just consider region III ({\it i.e.} $p_o>0$ and $r_o+l_o>0$) 
and multiply the result by an
extra factor $2$. Hence, the contribution of bremsstrahlung is given by
\begin{Eqnarray}
  &&{\rm Im}\,\Pi^{^{AR}}{}_\mu{}^\mu(q_o,{\imb 0})\approx
  {{NC_{_{F}}}\over{(2\pi)^4}} e^2g^2\int dp\int ldl \int dl_o
  \nonumber\\
  &&\qquad\times\;
  \left[n_{_{F}}(r_o)-n_{_{F}}(p_o)\right]
  \left[n_{_{B}}(l_o)+n_{_{F}}(r_o+l_o)\right]
  \nonumber\\
  &&\qquad\times
  \sum\limits_{a=T,L}\rho_a(L) P^a_{\rho\sigma}(L)
    \left[{{{\rm Trace}{}^{\rho\sigma}{}_{|_{\rm vertex}}}
        \over{\overline{R}{}^2(\overline{P+L}){}^2}}
      +{{{\rm Trace}{}^{\rho\sigma}{}_{|_{\rm self}}}
        \over{\overline{R}{}^2
          \overline{R}{}^2}}\right]_{{\rm sym\ \ \ \ }\atop{{p_o=p\ \ \ }
    \atop{\cos\theta^\prime}=\cdots}}\!\!\!.
\end{Eqnarray}

\subsection{Extraction of the dominant terms}
\subsubsection{General considerations}
Let us now concentrate on the matrix element to be integrated
over the phase space depicted above.
In order to perform the contraction of the Dirac's trace obtained in
Eq.~(\ref{eq:tracetotal}) with the longitudinal projectors, we use the
following expressions \cite{Weldo1,LandsW1}
\begin{eqnarray}
  &&P^{^{T}}_{\rho\sigma}(L)=
  g_{\rho\sigma}-U_\rho U_\sigma
  +{{(L_\rho-l_oU_\rho)(L_\sigma-l_oU_\sigma)}\over{l^2}}\\
  &&P^{^{L}}_{\rho\sigma}(L)=
  -P^{^{T}}_{\rho\sigma}(L)+g_{\rho\sigma}-{{L_\rho L_\sigma}\over{L^2}}
  \; ,
  \label{eq:projectors}
\end{eqnarray}
where $U\equiv(1,{\imb 0})$ is the 4-velocity of the plasma in its rest
frame.
An important simplification is obtained in the case of static photons
since we have
\begin{equation}
  Q^\rho P^{^{T}}_{\rho\sigma}=0\; .
\end{equation}
Taking into account this simplification, we
obtain after some algebra:
\begin{Eqnarray}
  &&\sum\limits_{a=T,L}\rho_a(L) P^a_{\rho\sigma}(L)
  \left[{{{\rm Trace}{}^{\rho\sigma}{}_{|_{\rm vertex}}}
      \over{\overline{R}{}^2(\overline{P+L}){}^2}}
    +{{{\rm Trace}{}^{\rho\sigma}{}_{|_{\rm self}}}
      \over{\overline{R}{}^2\overline{R}{}^2}}\right]_{\rm sym}
  \approx\nonumber\\
  &&\qquad-4\left[\left(\rho_{_{T}}(L)-\rho_{_{L}}(L)\right)
    {{4p^2(\cos^2\theta^\prime-1)}\over{R^2(P+L)^2}}
    \left(L^2-2{{Q^2(Q\cdot L)^2}\over{R^2(P+L)^2}}\right)\right.\nonumber\\
  &&\qquad+2{{(Q+L)^2}\over{R^2(P+L)^2}}
  \left(Q^2\rho_{_{L}}(L)+L^2\rho_{_{T}}(L)\right)\nonumber\\
  &&\qquad-2\rho_{_{T}}(L)\left(1-2{{(Q\cdot L)^2}\over{R^2(P+L)^2}}
      \right.\nonumber\\
      &&\qquad\qquad\qquad\left.\left.+{{Q^2L^2}\over 2}
          \left[{1\over{(R^2)^2}}+{1\over{((P+L)^2)^2}}
      \right]\right)
      \right]\; .
  \label{eq:static1}
\end{Eqnarray}
It is worth noticing that all the potential poles in $1/L^2$ (see
Eq.~(\ref{eq:projectors}-b)) have disappeared in this formula. This is
nothing but a consequence of the gauge invariance of the set of diagrams
we are looking at.

Now, in order to  extract the order of magnitude of each term in
Eq.~(\ref{eq:static1}), we can use the following very rough rules:

\noindent $\bullet$ $R^2\sim (P+L)^2 \sim 2p q_o$ 
which is correct for our purpose even if $l$ is
hard since $l$ is bounded by a quantity proportional to $p$ (see the
previous paragraph on kinematics).

\noindent $\bullet$ $n_{_{B}}(l_o)\sim T/l_o$ which 
always gives the correct order of
magnitude, even when $l_o$ is hard. 

\noindent $\bullet$ $n_{_{F}}(r_o+l_o)={\cal O}(1)$.

\noindent $\bullet$ 
$n_{_{F}}(r_o)-n_{_{F}}(p_o)\approx q_on^\prime_{_{F}}(p)$.

 \noindent $\bullet$ $1-\cos^2\theta^\prime={\cal O}(1)$.
 
\noindent $\bullet$ The details of the 
$x$ dependence are irrelevant 
to obtain the correct
order of magnitude.

\noindent $\bullet$ $\rho_{_{T,L}}(l,x)$ behaves 
like $m_{\rm g}^2/(l^2+m_{\rm g}^2)^2$ if one neglects its $x$ dependence.

Moreover, using the variable $x\equiv l_o/l$, we can
see ${\rm Im}\,\Pi^{^{AR}}{}_\mu{}^\mu(q_o,{\imb 0})$ is a sum of terms
like
\begin{equation}
  e^2g^2 q_o^\alpha\int dp\,n^\prime_{_{F}}(p) p^\beta \int dl\, l^\gamma
  \int dx\,
  F(x) \rho_{_{T,L}}(l,x) \left({{T}\over{l}}+{\cal
      O}(1)\right)\; ,
\end{equation}
where $F(x)$ is a dimensionless function and where
$\alpha+\beta+\gamma=3$ in order to give the correct overall dimension.
A close inspection of Eq.~(\ref{eq:static1}) shows that $\alpha$ ranges
from $\alpha=-1$ to $\alpha=3$, taking all the integer values between
these bounds. Eq.~(\ref{eq:static1}) shows also that $\beta\le 0$.
From the previous structure
and since we can factorize a factor $m_{\rm g}^2$ out of the spectral
functions $\rho_{_{T,L}}$, it is obvious that we can write the result
as
\begin{equation}
  {\rm Im}\,\Pi^{^{AR}}{}_\mu{}^\mu(q_o,{\imb 0})\approx e^2 g^2 
  {{m^2_{\rm g}T}\over{q_o}}P\left({{q_o}\over{m_{\rm g}}},
    {{q_o}\over{T}}\right)\; ,
\end{equation}
where $P(\cdot)$ is a polynomial of two variables, of total degree $4$.
Under the assumption that $q_o\ll T$,  we are allowed to truncate this
polynomial and keep only terms of degree $0$ in the variable $q_o/T$,
leaving a polynomial of $q_o/m_{\rm g}$ only. Moreover, the above rules
show that the term of Eq.~(\ref{eq:static1}) proportional to 
$p^2(\cos^2\theta^\prime -1)L^2$ contributes to the constant term of
this polynomial, and gives an integral that 
behaves like $\int dl/l$ for hard $l$.
This means that a logarithm of order $\ln(1/g)$
shows up in this coefficient. The argument of this logarithm can be
written as
$T^2/f(m^2_{\rm g},q_o^2)$ where $f(\cdot)$ is a function of dimension
two. This function depends on both $q_o$ and $m_{\rm g}$ 
since there can be a competition between
$q_o$ which appears as a kinematical infrared 
cut-off in the integral over $dl$
and $m_{\rm g}$ which appears in $\rho_{_{T,L}}$ and can also play the
role of an infrared cut-off for the same integral. Using the same tools,
it is quite easy to
check that all the other coefficients of this polynomial are of order
$1$ ({\it i.e.} do not contain any large logarithm).

Therefore, under the assumption that $q_o,m_{\rm g} \ll T$,
we can formally put the result into the compact form:
\begin{equation}
  {\rm Im}\,\Pi^{^{AR}}{}_\mu{}^\mu(q_o,{\imb 0})\approx C e^2 g^2 
  {{m^2_{\rm
        g}T}\over{q_o}}\left[
    \ln\left({{T^2}\over{f(m^2_{\rm g},q^2_o)}}
    \right)+Q\left({{q_o}\over{m_{\rm g}}}\right)\right]\; ,
  \label{generic}
\end{equation}
where $C$ is a numerical constant and $Q(\cdot)$ is a polynomial.

\subsubsection{Extraction of the logarithmic behavior}

Under some more restrictive assumptions, we can go further analytically.
More precisely, it is possible to extract analytically the constant $C$ 
in front of the logarithm of
Eq.~(\ref{generic}), as well as the function $f(m^2_{\rm g},q_o^2)$ 
in the
limit $q_o\ll m_{\rm g}$. The assumption $q_o,m_{\rm g}\ll T$ ({\it i.e.}
$g\ll 1$ and $q_o\ll T$) 
ensures that the argument of the logarithm is large, so that
the logarithmic term should be a fairly good approximation of the whole
expression\footnote{To summarize, the condition $q_o, m_{\rm g} 
\ll T$ is
essential to have a large logarithm, whereas the extra inequality
$q_o\ll m_{\rm g}$ is necessary just to be able to calculate 
analytically the function $f(m^2_{\rm g},q^2_o)$.}.

As mentioned before, the logarithm we are looking at comes from the term in 
$p^2(\cos^2\theta^\prime -1)L^2$ in Eq.~(\ref{eq:static1}). 
Using Eq.~(\ref{eq:cosine}), we obtain the following expression for
the imaginary part of the photon polarization tensor:
\begin{Eqnarray}
  &&{\rm Im}\,\Pi^{^{AR}}{}_\mu{}^\mu(q_o,{\imb 0})\approx
  -{{4NC_{_{F}}e^2g^2}\over{(2\pi)^4}} {T\over{q_o}}
  \int\limits_{0}^{+\infty}dp\,
  n^\prime_{_{F}}(p)\;
  \int\limits_{-1}^{+1}{{dx}\over{x}} (1-x^2)^2\nonumber\\
  &&\qquad\times\int\limits_{{q_o}\over{1-x}}^{{2p}\over{1-x}}
  dl\,l^3
  (\rho_{_{T}}(l,x)-\rho_{_{L}}(l,x))\; ,
\end{Eqnarray}
where
\begin{equation}
  \rho_{_{T,L}}(l,x)\equiv{{-2\,{\rm Im}\,\Pi_{_{T,L}}(x)}
    \over{(l^2(1-x^2)+{\rm Re}\,\Pi_{_{T,L}}(x))^2+({\rm
        Im}\,\Pi_{_{T,L}}(x))^2}}
\end{equation}
is the spectral function in the space-like region. This is the place
where the condition $q_o\ll m_{\rm g}$ enters the picture. 
Indeed, if we
don't make this assumption, the infrared regulator of the integral over
$dl$ will be a complicated combination of $q_o$ contained in the lower
bound and $m_{\rm g}$ contained in the spectral functions. On the
contrary, when $q_o\ll m_{\rm g}$, only the largest regulator 
({\it i.e.~}$m_{\rm g}$) plays a
role, and the argument of the logarithm is quite simple.
The integral over $dl$ is
elementary and yields an $\arctan$ function and a logarithm. 
Keeping only the latter\footnote{The $\arctan$ term we discarded is 
convergent when
performing the subsequent integration over $dx$ since the $\arctan$ is
bounded by $\pi/2$.}, we obtain:
\begin{Eqnarray}
  &&{\rm Im}\,\Pi^{^{AR}}{}_\mu{}^\mu(q_o,{\imb 0})\approx
  {{NC_{_{F}}e^2g^2}\over{(2\pi)^4}} {T\over{q_o}}
  \int\limits_{0}^{+\infty}dp\,
  n^\prime_{_{F}}(p)
  \int\limits_{-1}^{+1}{{dx}\over{x}} 
  \nonumber\\
  &&\quad\quad\times\left({\rm Im}\,\Pi_{_{L}}(x)-{\rm
      Im}\,\Pi_{_{T}}(x)\right)\ln\left({{p^4}\over{m^4_{\rm
          g}}}\right)\; ,
\end{Eqnarray}
The terms
neglected in that procedure show up only in the polynomial that
would accompany the $\ln(1/g)$ in a more complete calculation,
and are not tractable analytically. Moreover, for the same reason and
because of the statistical weight in the $dp$ integral that will cut off
everything above $p\sim T$, we can replace the remaining
logarithm\footnote{It is possible to perform analytically the $dp$
integral without this further approximation. Doing so leads to a result 
in which the logarithm of Eq.~(\ref{eq:staticfinal}) is replaced by
$\ln(T^2/m^2_{\rm g})+2(\ln(\pi/2)-\gamma)$ where $\gamma$ is the Euler
constant.
However, the additional 
constant is not complete since we have already neglected
contributions to it in earlier approximations.} by
$\ln(T^4/m^4_{\rm g})$. Now, the $dx$ and $dp$ integrals are trivial and
give:
\begin{equation}
  {\rm Im}\,\Pi^{^{AR}}{}_\mu{}^\mu(q_o,{\imb 0})\approx
  {{3NC_{_{F}}e^2g^2}\over{8\pi^3}} {{m^2_{\rm g} T}\over {q_o}}
  \ln\left({{T^2}\over{m^2_{\rm g}}}\right)\; .
\end{equation}
The production rate is then given by (see Eq.~(\ref{virtualphot})):
\begin{equation}
  \left.{{dN}\over{dtd{\imb x}}}\right|_{\rm bremss}\approx
  {{dq_od{\imb q}}\over{8\pi^6}}NC_{_{F}}
  \alpha^2\Big(\sum_{f}e^2_f\Big)
  \left({{m_{\rm g}\over {q_o}}}\right)^2
  \left({{gT}\over {q_o}}\right)^2 \ln\left({{T^2}
      \over{m^2_{\rm g}}}\right)\; ,
  \label{eq:staticfinal}
\end{equation}
where
the sum runs over the flavor of the quarks in the loop ($e_f$ is
the electric charge of the quark of flavor $f$, in units of the electron
electric charge).

A comment is relevant concerning the sensitivity of the exchanged gluon to 
the hard
scale. Indeed, the discontinuity of the effective propagator is used here 
in the
space-like region, and the HTL approximation used to obtain this propagator may
inaccurately reflect the phenomenon of Landau damping for a hard gluon. The
consequence of this remark is that a loop correction on the gluon propagator 
may lead to an important three-loop correction to the photon emission-rate.

\begin{figure}[t]
  \centerline{\rotatebox{-90}{\resizebox*{5cm}{!}{\includegraphics{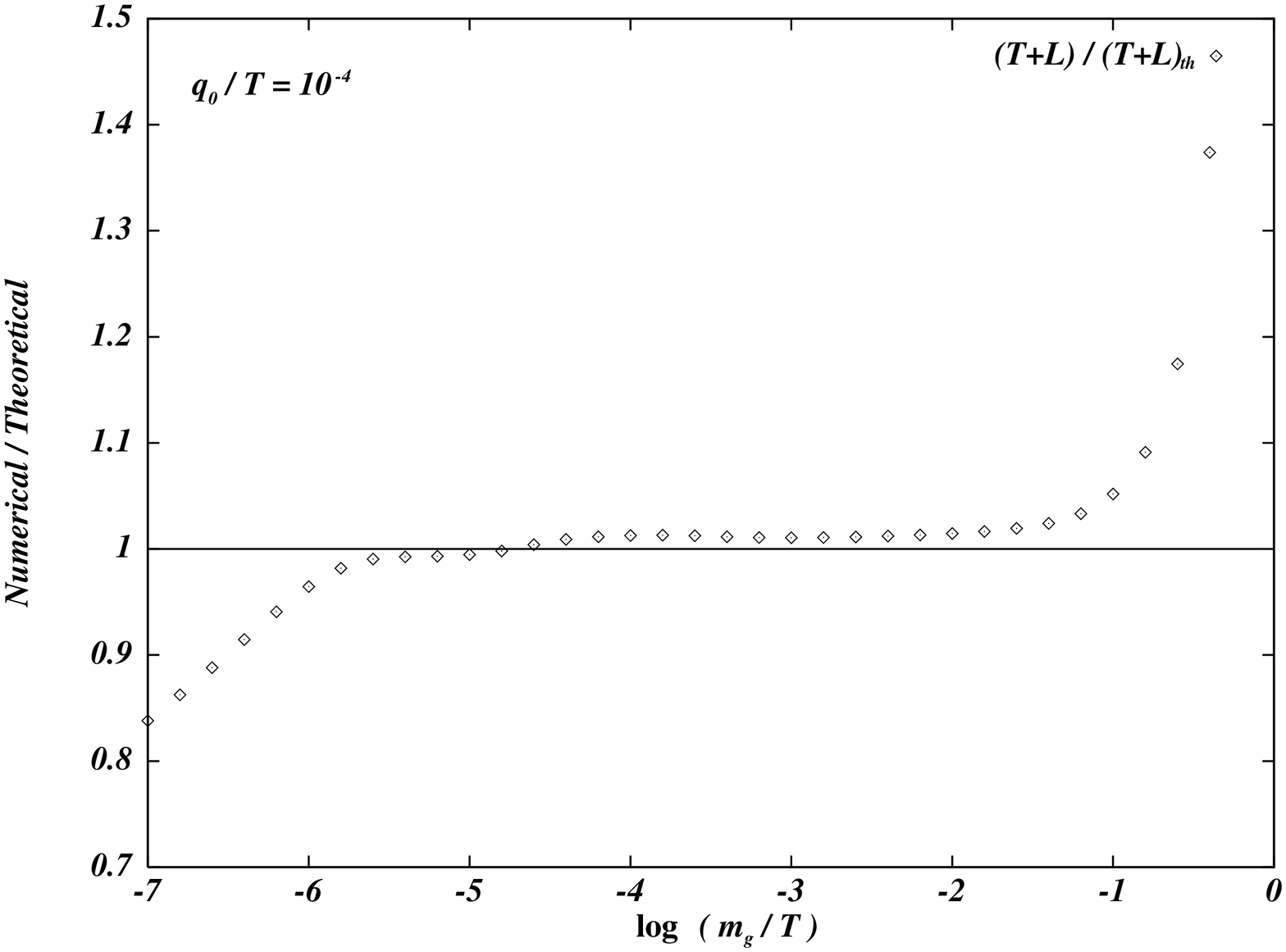}}}
    \rotatebox{-90}{\resizebox*{5cm}{!}{\includegraphics{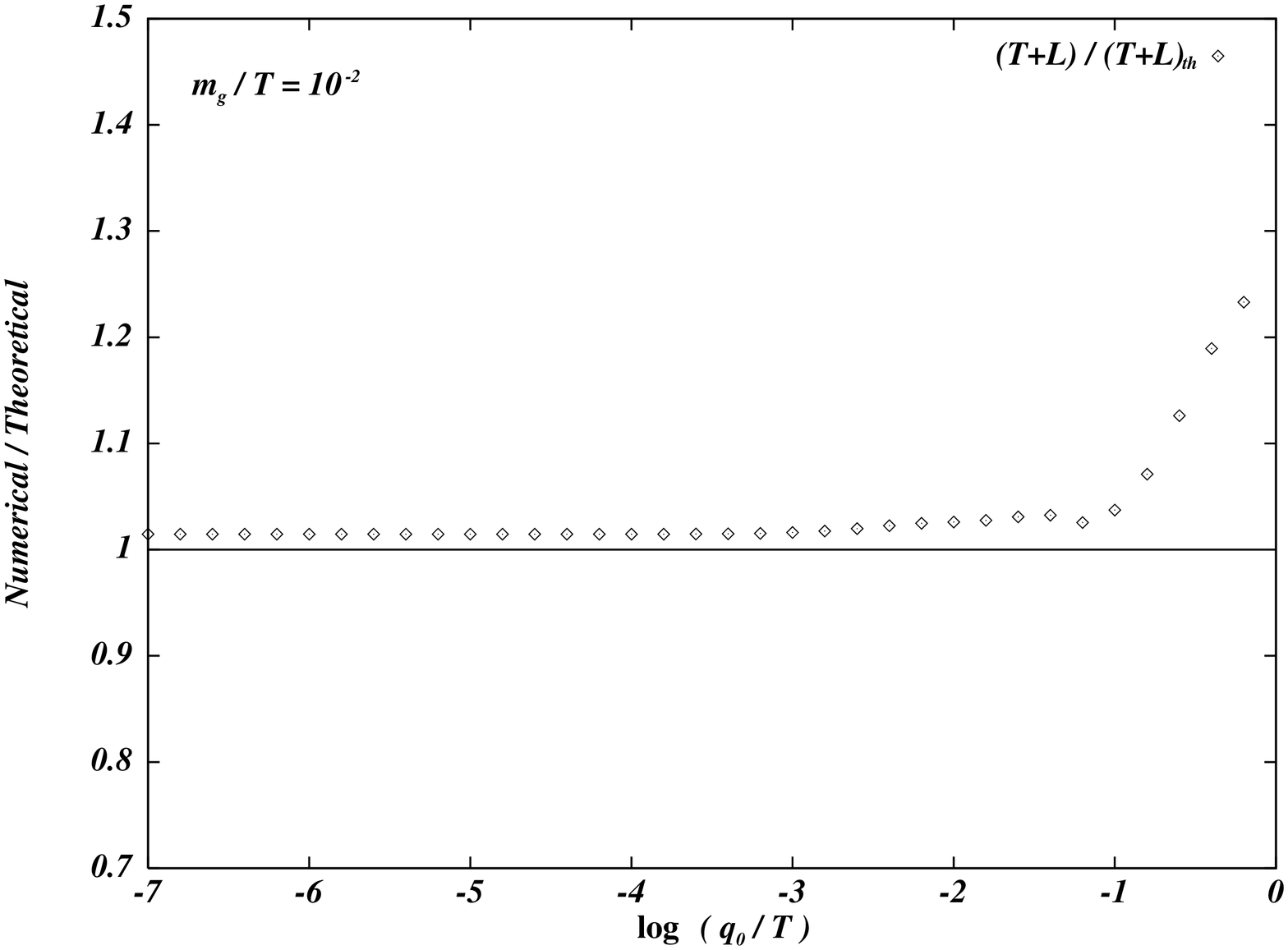}}}}
  \caption{\footnotesize{Comparison of numerical estimates of the complete
      matrix element with the simple theoretical expression obtained in
      Eq.~(\ref{eq:staticfinal}). Both plots show the ratio
      ``Numerical/Theoretical''. On the left plot, $q_o/T$ is fixed at
      $10^{-4}$ and we look at the variations with $m_{\rm g}/T$ ({\it i.e.~}
      with $g$). On the right plot, $m_{\rm g}/T$ is fixed at $10^{-2}$ and
      the photon energy varies from ultra-soft energies to hard ones.}}
  \label{fig:plotstatic}
\end{figure}

Before comparing this analytical result with numerical estimates of the
unapproximated expression, let us recall the domain in which this
expression is expected to be valid. Firstly, we need the logarithm to be
large in order to be dominant, which requires $q_o,m_{\rm g}\ll T$, 
{\it i.e.~}$g\ll 1$ and $q_o\ll T$. The additional {\it purely technical} 
condition is that $q_o\ll m_{\rm g}$, in order to keep simple the
argument of the logarithm. On the plots of the figure 
\ref{fig:plotstatic},
we show the ratio ``numerical/theoretical", where ``theoretical"
is the formula given in Eq.~(\ref{eq:staticfinal}) 
while ``numerical" denotes
a numerical evaluation of the contribution to bremsstrahlung of the
complete matrix element as given in Eq.~(\ref{eq:static1}).
The left plot shows that $m_{\rm g}/T$ must be smaller than $0.1$ in
order to have an agreement between our approximations and the complete
expression with an accuracy better than 5\%. If $m_{\rm g}/T$ is not
small enough, then the polynomial that comes with the $\ln(1/g)$ cannot
be neglected anymore. On the second plot, we see that 
the approximations we
performed inside the logarithm by assuming the smallness of $q_o/m_{\rm
g}$ are in fact valid far outside their expected domain of
validity\footnote{This is presumably due to the fact that this extra
assumption affects only the terms inside a slowly varying logarithm.},
since we still have a reasonable accuracy with $q_o/m_{\rm g}\sim 10$.

\subsection{Comparison with other approaches}
\subsubsection{Extrapolation of the quasi-real soft photons results}
In a previous paper, we gave asymptotic formulae for the same quantity
in the case where the photon invariant mass satisfies $M^2_\infty/T^2\ll
Q^2/q_o^2\ll 1$ (see Eqs.~(89),(90) and (93) of \cite{AurenGKP2}).
By extrapolating these estimates outside of their apparent range of
validity towards the case of static photons for which $Q^2/q_o^2=1$, we
obtain exactly the formula of Eq.~(\ref{eq:staticfinal}). Such an
agreement means that, for a given energy $q_o$, 
the formulae established for the production of low
invariant mass photons are very robust since they 
remain valid when
extrapolated to the case of heavy photons at rest. 

From a technical point of
view, this is made possible by the fact that the term that contains the
collinear singularity in which we were interested for low mass photons
and the term that develops the logarithm we extracted
analytically for heavy photons are the same.

\subsubsection{Braaten et al. results}
\label{BPYcomp}
At the one loop level in the effective perturbative expansion, 
the production rate of soft static photons has been evaluated by Braaten,
Pisarski and Yuan in
\cite{BraatPY1}. The purpose of this paragraph is to present an analytic
comparison of their final result (Eq.~(11) of \cite{BraatPY1}) with
Eq.~(\ref{eq:staticfinal}), in the domain $q_o\ll m_{\rm g}\sim gT\ll T$
for which our expression has been justified. In this domain, we can
retain from the result of \cite{BraatPY1} only the terms having the most
singular behavior in $q_o$. Such terms are found only in the ``cut-cut''
part. Moreover, some of these terms develop a large logarithm $\ln(1/g)$
which is simple to extract analytically, in a way very similar to the
method leading to 
Eq.~(\ref{eq:staticfinal}). Applying these approximations
to BPY's result leads to the following estimate for the one-loop
production rate:
\begin{equation}
  \left.{{dN}\over{dtd{\imb x}}}\right|_{\rm 1-loop}\approx
  {{dq_od{\imb q}}\over{12\pi^4}}N\alpha^2\Big(\sum_{f}e^2_f\Big)
  \left({{m_{_{F}}\over {q_o}}}\right)^4 \ln\left({{T^2}
      \over{m^2_{_{F}}}}\right)\; ,
\end{equation}
where $m^2_{_{F}}=M^2_\infty/2$ is the soft quark thermal mass.
Therefore, comparing with Eq.~(\ref{eq:staticfinal}),
we obtain the ratio:
\begin{equation}
  {{\left.{{dN}}\right|_{\rm bremss}}
    \over{\left.{{dN}}\right|_{\rm 1-loop}}}\approx
  {{32}\over{3\pi^2}}{{N+N_{_{F}}/2}\over{C_{_{F}}}}\; ,
\end{equation}
which for 2 light flavors and 3 colors becomes
\begin{equation}
  {{\left.{{dN}}\right|_{\rm bremss}}
    \over{\left.{{dN}}\right|_{\rm 1-loop}}}\approx
  {{32}\over{\pi^2}}\sim 3.2\; .
\end{equation}
This ratio is rather large, which means that bremsstrahlung is definitely
an essential contribution to the soft static photon production rate by a hot
plasma.

\subsubsection{Cleymans et al. results}
The bremsstrahlung production of a soft virtual photon has been
considered in the context of the semi-classical approximation by
Cleymans {\em et al.}. In their approach\cite{CleymGR2}, 
they took into account the
effect of the multiple scattering of the quark in the plasma
(Landau-Pomeranchuck-Migdal effect). In order to compare with our
thermal field theory result we need to ``undo'' the effect of
rescattering and consider only one collision of the (photon emitting)
quark in the plasma. Cleymans {\em et al.} use several simplifying
hypotheses: the energy of quarks or gluons is much larger than the
temperature so that Boltzmann distributions are used for particles
entering the interaction region and a factor 1 is assigned to those
leaving it. The scattering of quark in the plasma is treated as in the
vacuum, the only modification being the introduction of a
phenomenological Debye mass $m_{_D}$ to screen the forward singularity
of the quark scattering amplitude. Neglecting
furthermore the virtual photon momentum compared to the momenta of the
constituents in the plasma the production rate can be factorized into a
quark scattering term and a photon emission term so that the lepton pair
rate can be written (see \cite{AurenGKP2} for a similar expression in the
case of a real photon)
\begin{Eqnarray}
  \openup 15mm
  \displaystyle{{{dN}\over{dtd{\imb x}}}}
  &&\approx {d q_o d{\imb q}\over{24\pi^4 }}{\alpha\over Q^2}
  \int\prod\limits_{i=1,2} e^{-p_i/T}
  {d{\imb p}_i \over{(2\pi)^3} 
    2 p_i}
  \prod\limits_{i=1,2}
  {d{\imb p'}_i \over{(2\pi)^3} 
    2 p'_i}
  \nonumber\\
  &&\times
  (2\pi)^4 \delta(P_1+P_2-P'_1-P'_2)
  \nonumber\\
  &&\times |{\cal M}|^2(P_1,P_2;P'_1;P'_2)
  \ e^2\sum\limits_{\hbox{\rm pol.}\ \varepsilon}\left(
    {{P_1\cdot\varepsilon}\over{P_1\cdot Q}}-
    {{P'_1\cdot\varepsilon}\over{P'_1\cdot Q}}
  \right)^2,
  \label{factorization}
\end{Eqnarray}
where $|{\cal M}|^2$ is the square of the matrix element of the quark
scattering process. We have folded in the above expression the
appropriate factor describing the decay of the virtual photon of mass
$Q$ into the lepton pair. With the above mentioned approximations and
keeping the most singular term in the $t-$channel \cite{CleymGR2} we
have
\begin{equation}
  |{\cal M}|^2(P_1,P_2;P'_1,P'_2) = C_{ab}\ 2 g^4 \ 
  \left({2 P_1 \cdot P_2 \over 2 P_1 \cdot P'_1}\right)^2
\end{equation} 
with $C_{ab}=4/9$ for quark-quark scattering and
$C_{ab}=1$  for quark-gluon scattering. We then find for the rate of
production of the pair at rest the following expression:
\begin{equation}
  \displaystyle{{{dN}\over{dtd{\imb x}}}} \approx 
  {d q_o d{\imb q}\over{3\pi^6 }}{\alpha^2\alpha_s^2}{d_{f} d}
  \Big(\sum_{f}e^2_f\Big){T^4\over q_o^4} \ln\left({{T^2}
      \over{m^2_{_{D}}}}\right)\; ,
\end{equation}
where $d_f=2_s\times 3_c = 6$ and 
$d=\frac{4}{9}\ 2\times 2_f\times 2_s\times 3_c + 2_s\times 8_c 
= 26+{2\over 3}$ 
are the degeneneracy factors introduced
in \cite{CleymGR2}. Comparing with Eq.~(\ref{eq:staticfinal}) we find for
two light flavors
\begin{equation}
  {\left.{dN}\right|_{\rm semi-class} \over \left.{dN}\right|_{\rm bremss}}
  \approx {15 \over \pi^2}\; ,
\end{equation}
{\em i.e.~}the semi-classical result agrees with the thermal field
theory result in its functional dependence but over-estimates the rate
of production by about $50 \%$. This difference appears to be due to the
very approximate treatment  of thermal effects of the dynamics of the
plasma: for example, the ratio of quark-gluon scattering to quark-quark
scattering is estimated to be $N/ .5 N_{_F}= 3$ in thermal field theory
(this is the ratio of the gluon contribution to the quark contribution
in  a hard thermal loop) compared to $1.5$ (taking account of the
multiplicity factors associated to the quarks and the gluons) in the
semi-classical approach.

\section{Contribution to hard real photons}
\subsection{Kinematics}
Let us now concentrate on the case of hard real photons ($T\lesssim q_o=q$). 
The kinematics for this situation is much more
complicated than for static soft photons for two reasons. First, since
$Q$ is hard, we are no longer allowed to neglect $Q$ in front of $P$ or
$R$. Moreover, since we are looking this time at real photons, we may
encounter collinear divergences (like in the case of soft real photons
\cite{AurenGKP2}), and we must carefully keep the quark asymptotic 
thermal mass $M_\infty$ in the expressions.

Now, from the identity
$S^{^{R}}(P)-S^{^{A}}(P)=2\pi\epsilon(p_o)\delta(P^2 - M^2_\infty)$, we
extract the value $p_o=\pm\omega_p$, where we denote
$\omega_p\equiv\surd{({\imb p}^2+M^2_\infty)}$ and $r_o=q_o\pm\omega_p$.
The second delta function constraint provides us with the 
angle $\theta^\prime$ 
between the 3-vectors ${\imb r}$ and ${\imb l}$, via the relation:
\begin{equation}
  \cos\theta^\prime={{R^2-M^2_\infty+2r_ol_o+L^2}\over{2rl}}\; .
\end{equation}
Again, we must enforce the requirement $-1\le\cos\theta^\prime\le 1$,
which implies the following set of inequalities:
\begin{eqnarray}
  &&(r_o-r+l_o+l)(r_o+r+l_o-l)\ge M^2_\infty\\
  &&(r_o-r+l_o-l)(r_o+r+l_o+l)\le M^2_\infty\; ,
  \label{hardconstraints}
\end{eqnarray}
and leads to a reduction of the allowed domain in the $(l_o,l)$ plane.
\begin{figure}[ht]
  \centerline{\resizebox*{!}{5cm}{\includegraphics{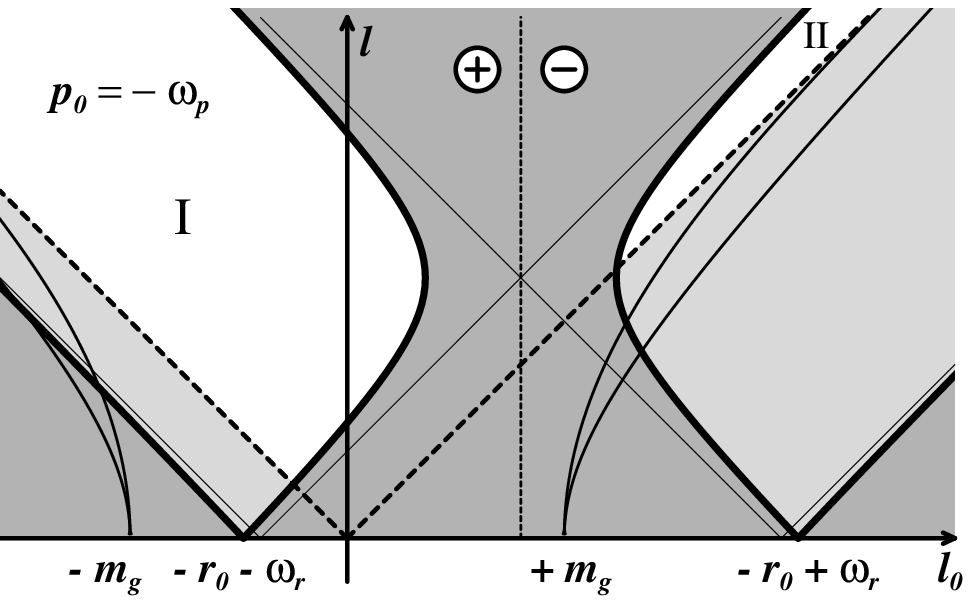}}}
  \centerline{\resizebox*{!}{5cm}{\includegraphics{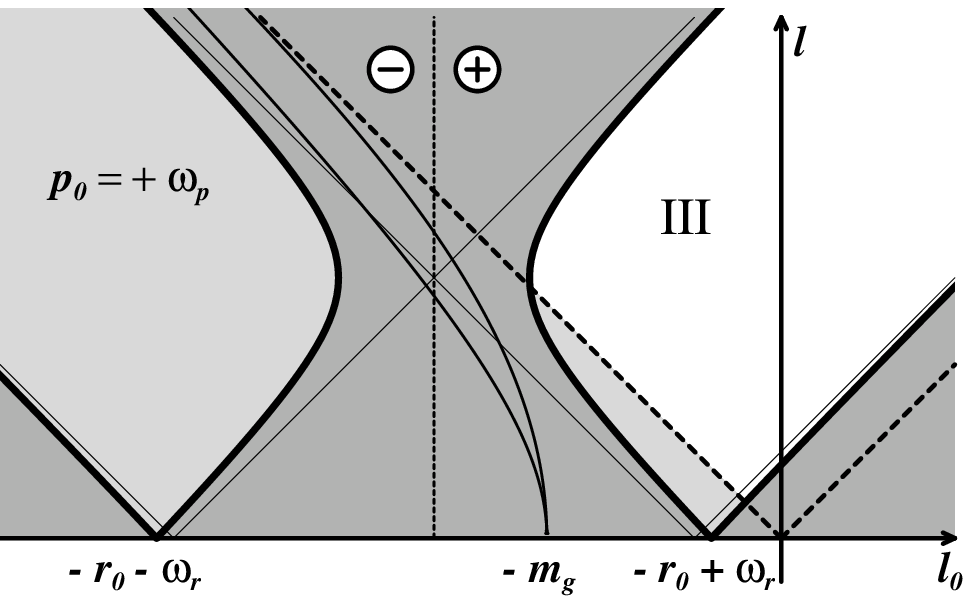}}}
  \caption{\footnotesize{Allowed domains in the $(l_o,l)$ plane for 
      $p_o=\pm \omega_p$. 
      The area shaded in dark
      gray is excluded by the delta function constraints.
      The areas shaded in light gray are above the light-cone (dotted lines).
      The light curves are the transverse and
      longitudinal mass shells of the thermalized gluon. 
      The vertical dotted
      line is the separation between $\epsilon(p_o)\epsilon(r_o+l_o)=+1$ and 
      $\epsilon(p_o)\epsilon(r_o+l_o)=-1$.}}
  \label{fig:kinehard}
\end{figure}
The above two inequalities may be rewritten as
\begin{eqnarray}
  &&\sqrt{(r-l)^2+M^2_\infty}\le l_o+r_o{\rm \ \ or \ \ }
  l_o+r_o \le -\sqrt{(r-l)^2+M^2_\infty}\qquad\qquad\\
  &&-\sqrt{(r+l)^2+M^2_\infty}\le l_o+r_o \le \sqrt{(r+l)^2+M^2_\infty}
  \label{eq:hardconstraints2}
\end{eqnarray}
The result of these inequalities is shown in Fig.~\ref{fig:kinehard},
where the regions excluded by the requirement
$\cos\theta^\prime\in[-1,+1]$ are shaded in dark gray. In order to make
the comparison easier with the case of static photons, we have also
reproduced the boundaries of the allowed region for that case 
({\it i.e.~}the frontiers that one obtains in the 
limit $M_\infty\to 0$). One can
see that the old boundaries are asymptotes for the news ones.
Again, we have three regions allowed by the above two
inequalities for $L^2\le 0$. Again, regions I and III give the same
contribution, as can be seen by the change of variables $P\to -R-L$
performed at an earlier stage of the calculation. From now on, we will
drop region I and multiply by a factor $2$ the contribution of region III.
We start the discussion with the study of region III (bremsstrahlung processes)
and then turn to region II ($q\bar{q}$ annihilation with scattering in the
plasma, see Fig.~\ref{fig:processes}). As will be seen, this region can no
longer be neglected, contrary to the case of soft photon production.

\subsection{Bremsstrahlung}
Since the delta function $\delta(P^2-M^2_\infty)$ makes more convenient
the choice of $p\equiv||{\imb p}||$ as an independent variable (instead
of $r$), the quantities $-r_o\pm\omega_r$ that appear on the previous
figure are to be understood as functions of $p$ and the angle $\theta$
between ${\imb p}$ and ${\imb q}$. Therefore, this restriction of the
allowed phase space is in fact a constraint that relates the independent
variables $p$,$\theta$,$l$ and $l_o$. 
In the following, doing the integration over $\theta$ first will prove
to be convenient since, due to collinear divergences, the result of this
angular integral controls the order of magnitude of the result. 
If $u\equiv 1-\cos\theta$ is the variable over which we integrate first,
then the bounds on $u$ depend on the other variables $l,l_o$ and $p$
because of the identities of Eqs.~(\ref{eq:hardconstraints2}-a) and
(\ref{eq:hardconstraints2}-b). 
In particular, the inequality Eq.~(\ref{eq:hardconstraints2}-a) can be
written in terms of $u\equiv 1-\cos\theta$:
\begin{equation}
  u\le {{(p+q)^2-(\sqrt{(r_o+l_o)^2-M^2_\infty}-l)^2}\over{2pq}}\; ,
\end{equation}
which gives if one assumes that $l_o$ and $l$ are soft
\begin{equation}
  u\le \left( {{l}\over{p}}+{{l}\over{q}}\right)(1-x)\; ,
\end{equation}
where we denote $x\equiv l_o/l$. As a consequence, since the relevant
values of the momentum $l$ are controlled by thermal masses of order
$gT$, this upper bound is of order $g$.

It is now worth giving
expressions for the denominators that enter in the rate, since they
are potentially dangerous when the photon is emitted collinearly to the
quark. To that purpose, we 
need also another angular variable, which is left unconstrained by
the previous considerations. This variable can be the
azimuthal angle $\phi$
between ${\imb q}$ and ${\imb l}$ when projected on a
plane orthogonal to ${\imb r}$. Therefore, if we denote
$\theta^{\prime\prime}$ the angle between ${\imb q}$ and ${\imb l}$, we
can calculate this angle by
\begin{equation}
  \cos\theta^{\prime\prime}=\cos\theta_r\cos\theta^\prime 
  +\sin\theta_r\sin\theta^\prime\cos\phi\; ,
  \label{eq:theta2}
\end{equation}
where $\theta_r$ is the angle between ${\imb r}$ and ${\imb q}$. This
last angle is not independent of $\theta$, to which it can be related by
\begin{eqnarray}
  &&r\sin\theta_r=p\sin\theta\\
  &&r\cos\theta_r=p\cos\theta+q\; ;
\end{eqnarray} 
so that Eq.~(\ref{eq:theta2}) can be rewritten as
\begin{equation}
  \cos\theta^{\prime\prime}={p\over{r}}\left(
    \cos\theta\cos\theta^\prime+\sin\theta\sin\theta^\prime\cos\phi\right)
  +{q\over r}\cos\theta^\prime\; .
\end{equation}
Using only the variables $u=1-\cos\theta$, $p$, $l$ and $x=l_o/l$, we
can write\cite{AurenGKP2}
\begin{equation}
  R^2-M^2_\infty\approx 2pq\left[u+{{M^2_\infty}\over{2p^2}}\right]\; ,
  \label{eq:R2}
\end{equation}
and
\begin{equation}
  \int\limits_{0}^{2\pi}{{d\phi}\over{(P+L)^2-M^2_\infty}}\approx
  {{2\pi(p+q)}\over{2qp^2\left[
        \left(u+{{M^2_\infty}\over{2p^2}}+{{L^2}\over{2p^2}}\right)^2
        -{{L^2}\over{p^2}}{{M^2_\infty}\over{p^2}}\right]^{1/2}}}\; ,
\label{eq:PL2}
\end{equation}
where we performed the integration over $\phi$ at this stage since this
denominator is the only place where $\phi$ appears at the dominant
order. As one can see, the first formula remains exactly 
the same as in the case
of soft real photons \cite{AurenGKP2}, while the second one is only
slightly modified by an extra $(p+q)/p$ factor.
As a consequence, the discussion made in \cite{AurenGKP2} concerning
the enhancement that one gets when performing the $u$ integral is still
valid. In particular, the terms where both denominators are
present are enhanced by a factor of order $p^2/M^2_\infty$,
while those where only one of them appear will get only a logarithm of
this quantity.

Again, our starting point is Eq.~(\ref{eq:tracetotal}). The order of
magnitude of each term is evaluated by taking into account the fact that
the momentum $Q$ is now hard, as well as the quark momentum. Moreover,
we must take into account the enhancement by a factor of order $1/g^2$
for terms having two denominators. Since the emitted photon is assumed
to be real, we have $Q^2=0$. Therefore, one can check that only one term
dominates in this matrix element:
\begin{equation}
  -8L^2{{\overline{R}{}^\rho\overline{R}{}^\sigma+
      \overline{P}{}^\rho\overline{P}{}^\sigma}
    \over{\overline{R}{}^2(\overline{P+L}){}^2}}\; ,
\end{equation}
which is again the same as in the case of soft real photons.

Since the relevant values of $u$ are of order
$M^2_\infty/p^2\sim g^2$, we have
\begin{equation}
  r_o-\omega_r\approx
  {{pq}\over{p+q}}\left[u+{{M^2_\infty}\over{2p^2}}\right]{\rm\ \ for\ }
  {p_o=+\omega_p}\; .
  \label{eq:diff1}
\end{equation}
Therefore, one can see in Fig. \ref{fig:kinehard}
that the extra requirement $L^2\le 0$ 
does not restrict significantly region III due to
Eq.~(\ref{eq:diff1}) (this is
a result already obtained in \cite{AurenGKP2} 
according to which the kinematics confines
the phase space to the $L^2\le 0$ region in the collinear limit if $L$
is soft). 

Since the term to consider is the same, and since the denominators have
very similar expressions, it is completely straightforward to reproduce
the calculation\footnote{As seen earlier, the kinematics gives an upper
bound of order $g$ for the variable $u$. Since we are considering
enhanced angular integrals for which the relevant values of $u$ are of
order $M^2_\infty/p^2\sim g^2$, this upper bound does not appear in the
result of the $u$ integral.} that has been performed in 
\cite{AurenGKP2}.  The only minor difference lies in the contraction of
$\overline{R}{}^\rho\overline{R}{}^\sigma+
\overline{P}{}^\rho\overline{P}{}^\sigma$ with the projectors, which 
gives a factor $p^2+r^2$ approximated by
$p^2+(p+q_o)^2$ in the collinear limit. Introducing
the dimensionless quantities:
\begin{eqnarray}
  &&w\equiv{{-L^2}\over{M^2_\infty}}\\
  &&\widetilde{I}_{_{T,L}}(x)\equiv{{{\rm Im}\,\Pi_{_{T,L}}(x)}\over
    {M^2_\infty}}\; ,\qquad
  \widetilde{R}_{_{T,L}}(x)\equiv{{{\rm Re}\,\Pi_{_{T,L}}(x)}\over
    {M^2_\infty}}\; ,
\end{eqnarray}
we find
\begin{Eqnarray}
  &&{\rm Im}\,\Pi^{^{AR}}{}_\mu{}^\mu(Q)\approx {{e^2 g^2 N
      C_{_{F}}}\over{\pi^4}}(J_{_{T}}-J_{_{L}}) {{T}\over{q_o^2}}\;\nonumber\\
  &&\qquad\times\int\limits_{0}^{+\infty}dp\;
  (p^2+(p+q_o)^2)\left[n_{_{F}}(p)-n_{_{F}}(p+q_o)\right]\; ,
  \label{eq:hardresult}
\end{Eqnarray}
where we denote
\begin{equation}
  J_{_{T,L}}\equiv\int\limits_{0}^{1}{{dx}\over{x}}\;
  \widetilde{I}_{_{T,L}}(x)\;\int\limits_{0}^{+\infty}dw\;
  {{\sqrt{w/{w+4}}\,{\rm
        tanh}^{-1}\sqrt{w/{w+4}}}\over{(w+\widetilde{R}_{_{T,L}}(x))^2
      +(\widetilde{I}_{_{T,L}}(x))^2}}\; .
\end{equation}
As one can see, the integral over $x$ and $w$ can be factorized out of
the expression and is the same as in the case of soft real photons
\cite{AurenGKP2}. In this previous work, we noticed that the
integral over the gluon momentum could potentially be singular when the
gluon is transverse, but is regularized by the {\it quark} thermal mass
instead of the gluon thermal mass as one might expect. The
interpretation of this result is in fact rather simple. Indeed, a close
look at kinematics shows that the delta functions
$\delta(P^2-M^2_\infty)$ and $\delta((R+L)^2-M^2_\infty)$ become
incompatible in the limit $L\to 0$ if $M^2_\infty > 0$. Therefore, the
quark thermal mass prevents an infrared divergence by reducing to zero
the region of phase space where this divergence can occur.
The only major difference with \cite{AurenGKP2} concerns the
integral over $p$, because of the fact that $q_o$ is now hard.
Nevertheless, we can still give a rather compact expression for this
integral in terms of poly-logarithms:
\begin{Eqnarray}
  &&\int_0^\infty dp (p^2+(p+q_o)^2) [n_F(p) - n_F(p+q_0)] 
  \nonumber\\
  &&\qquad=
  T^3 \left[   3\zeta(3)
    +{{\pi^2}\over{6}} {{q_o}\over{T}}
    +\left({{q_o}\over{T}}\right)^2\ln(2) \right.\nonumber\\    
  &&\qquad \qquad+4\,{\rm Li\kern 1pt}_3(-e^{-|q_o|/T})
  +2 {{q_o}\over{T}}\, {\rm Li\kern 1pt}_2(-e^{-|q_o|/T})
  \nonumber\\
  &&\qquad\qquad\left.
    -\left({{q_o}\over{T}}\right)^2 \ln(1+e^{-|q_o|/T})  \right]\; ,
  \label{eq:polylogs}
\end{Eqnarray}
where the poly-logarithm functions are defined via
\begin{equation}
  {\rm Li\kern 1pt}_a( z)\equiv\sum\limits_{n=1}^{+\infty}
  {{z^n}\over{n^a}}\; .
\end{equation}
Eq.~(\ref{eq:polylogs}) simplifies in the limit of
extremely hard photons $q_o\gg T$:
\begin{equation}
  \int\limits_{0}^{+\infty}dp\;
  (p^2+(p+q_o)^2)\left[n_{_{F}}(p)-n_{_{F}}(p+q_o)\right]
  \build{\approx}\over{q_o\gg T}q_o^2 T \ln(2)\; .
\end{equation}
Therefore, in this asymptotic regime, we find for the production rate of
hard real photons:
\begin{equation}
  \left.{{dN}\over{dtd{\imb x}}}\right|_{\rm bremss}
  \build{\approx}\over{q_o\gg T}
  2{{d{\imb q}}\over{\pi^5}}NC_{_{F}}\alpha
  {{\alpha_{_{S}}}}
  \Big(\sum_{f} e_f^2\Big)
  {{T^2}\over{q_o}} e^{-q_o/T}(J_{_{T}}-J_{_{L}})\ln(2)\; ,
  \label{eq:hardfinal}
\end{equation}
where $e_f$ is the electric
charge of the quark of flavor $f$ expressed in units of the electron
charge.

Besides this simple asymptotic result, it is worth adding that
Eqs.~(\ref{eq:hardresult}) and (\ref{eq:polylogs}) provide a
generalization of the analogous formula of \cite{AurenGKP2}. Indeed, the
result we provide in the present paper for the bremsstrahlung
contribution to real photon production is valid over the whole range of
photon energies, and in particular reduces to Eq.~(55) of
\cite{AurenGKP2} in the limit of soft $q_o$.

\subsection{$q\bar{q}$ annihilation with scattering}
The discussion of the contribution of region II can be carried out in a similar
way. We recall that $p_o=-\omega_p< 0$ so that Eq.~(\ref{eq:R2}) becomes
\begin{equation}
R^2-M^2_\infty\approx-2pq\left[ v+{{M^2_\infty}\over{2p^2}}\right]\; ,
\end{equation}
with the notation $v\equiv 1+\cos\theta$. For hard enough $q_o$, the 
statistical
weight $n_{_{F}}(p_o)-n_{_{F}}(p_o+q_o)$ is equal to $1$ for 
$-q_o\le p_o \le 0$
and equals $0$ everywhere else\footnote{The transition between $0$ and 
$1$ takes
place in a range of width $T$ in the variable $p_o$. These side-effects are
negligeable if the condition $q_o\gg T$ is satisfied since they provide
corrections of relative order $T/q_o$.}. 
Therefore, we can restrict to the range $0\le
p \le q_o$ and we check that Eq.~(\ref{eq:PL2}) remains valid except
for the changes $u\to v$ and $q+p \to q-p$. The enhancement mechanism in the
terms that contain the two denominators goes through as before due to their
behavior near $v\sim g^2$. 

Such terms are dominated by the region where $v\sim g^2$, 
which enables us to obtain $-r_o-\omega_r \approx -2(q_o-p)$ and
$-r_o+\omega_r\sim{\cal O}(g^2T)$. As a consequence, the boundaries in the
$(l_o,l)$ plane are $-1\le x \le 1$ and $0\le l\le 2(q-p)/(1-x)$, where 
$x\equiv l_o/l$. Therefore, the integral over $l$ and $x$ gives the same factor
$J_{_{T}}-J_{_{L}}$ as before. The integral over $p$ in
Eq.~(\ref{eq:hardresult})
is now to be replaced by
\begin{equation}
\int\limits_{0}^{q_o}dp\,(p^2+(q_o-p)^2)={{2q_o^3}\over{3}}\; ,
\end{equation}
which leads to the following asymptotic contribution for the region II:
\begin{equation}
\left.{{dN}\over{dtd{\imb x}}}\right|_{\rm II}
  \build{\approx}\over{q_o\gg T}
  {2\over 3}{{d{\imb q}}\over{\pi^5}}NC_{_{F}}\alpha
  {{\alpha_{_{S}}}}
  \Big(\sum_{f} e_f^2\Big)
  {{q_oT}\over{q_o}} e^{-q_o/T}(J_{_{T}}-J_{_{L}})\; .
  \label{eq:hardIIfinal}
\end{equation}
Therefore, it appears that region II dominates over bremsstrahlung in the
asymptotic regime.

\subsection{Comparison with previous results}
The production rate of hard real photons has already been calculated at
the one-loop order in \cite{BaierNNR1,KapusLS1}
as an application
of the effective theory based on the resummation of hard thermal loops.
We now compare the contribution of bremsstrahlung
obtained above with their result. For
hard real photons, the predictions of \cite{BaierNNR1} are
\begin{equation}
  \left.{{dN}\over{dtd{\imb x}}}\right|_{\rm 1-loop}
  \build{\approx}\over{q_o\gg T}
  {{d{\imb q}}\over{8\pi^2}}NC_{_{F}}\alpha
  {{\alpha_{_{S}}}}
  \Big(\sum_{f} e_f^2\Big)
  {{T^2}\over{q_o}} e^{-q_o/T}\ln\left(
    {{cq_o}\over{\alpha_{_{S}}T}}\right)\; ,
  \label{eq:baierresult}
\end{equation}
with the constant $c\approx 0.23$.

It is worth recalling here that the quantities $J_{_{T,L}}$ that appear
in Eqs.~(\ref{eq:hardfinal}) and (\ref{eq:hardIIfinal}) are functions of
the ratio $M_{\infty}/m_{\rm g}$ when $Q^2=0$, {\em i.e.} 
depend only on $N$ and $N_{_{F}}$.  For $N=3$ colors and
$N_{_{F}}=2$ light flavors, we can evaluate numerically $J_{_{T}}\approx
4.45$ and $J_{_{L}}\approx -4.26$. The following plot shows a comparison
of bremsstrahlung, $q\bar{q}$ annihilation with scattering, and one-loop
contributions. 
\begin{figure}[ht]
  \centerline{\rotatebox{-90}{\resizebox*{6cm}{!}{\includegraphics{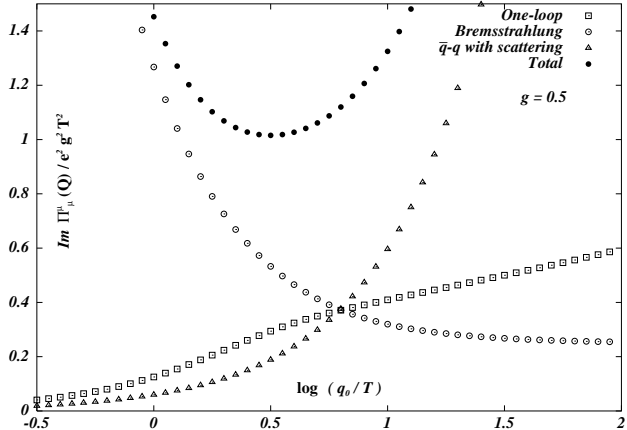}}}}
  \caption{\footnotesize{Comparison of various contributions to ${\rm
  Im}\,\Pi{}^\mu_\mu(Q)$ for a hard real photon. The comparison is made for
  $N=3$ colors and $N_{_{F}}=2$ flavors. The value taken for the coupling
  constant is $g=0.5$.}}
  \label{fig:comphard}
\end{figure}
In this figure, the bremsstrahlung contribution is taken from
Eq.~(\ref{eq:hardresult}); for the region II, we use the asymptotic result
obtained for large $q_o$ \footnote{This asymptotic estimate may be incorrect 
for
smaller values of $q_o$. Nevertheless, we expect it to decrease like $q_o$ 
since
its support in the $(l_o,l)$ plane decreases like $q_o$, so that region II is
certainly subdominant for $q_o\lesssim T$. Therefore, the
asymptotic result is sufficient in this comparison.}, 
and the one-loop result is a numerical evaluation of the
diagrams considered in \cite{BaierNNR1}. The value used here for the coupling
constant is $g=0.5$. We can see that the bremsstrahlung is the dominant
contribution for the smaller values $q_o$, whereas the region II becomes 
dominant for hard
enough $q_o$. For intermediate photon energies (around $T$), the three
contributions have equivalent orders of magnitude. For higher values of the
coupling constant $g$, the relative importance of the one-loop contribution 
tends to decrease. We can also see that the sum of the three contributions is
significantly above the one-loop contribution considered alone.

\section{Conclusions and perspectives}
We have discussed the production of a real photon or of a lepton pair in
a hot quark-gluon plasma in thermal equilibrium. We assumed that the plasma is
described by  thermal Quantum Chromodynamics and we worked in the
framework of the effective theory obtained after resummation of hard
thermal loops. We have shown that, at leading order in $g T$, it is
necessary to include the one-loop as well as the two-loop diagrams. The
one-loop diagrams of the effective theory correctly account for the
contribution of soft fermion momenta in the loop, whereas in the
two-loop diagrams hard momenta play a dominant role.

Many physical processes are contained in the effective theory calculated
up to two-loop order. To simplify, it can be said that, at one-loop,
scattering processes mediated by (soft) fermion exchanges are of
paramount importance for the emission of the photon. The corresponding
photon production rate consequently strongly depends on the soft fermion
thermal mass $m_{_F}$. On the other hand, the two-loop topologies account,
among other possibilities, for bremsstrahlung processes where the photon
is emitted by a hard quark scattering in the plasma via a (space-like)
gluon exchange: in that case the rate is proportional to the square of
the gluon mass $m_{\rm g}$.  We have calculated the contribution of such
processes. 

For the production a soft virtual photon ($q_o \sim g T$, ${\imb q} =
{\imb 0}$) the bremss\-tra\-hlung processes largely dominate over the
one-loop result in the range $q_o \lesssim m_{_F}$. For the case of real
photon production an interesting enhancement phenomenon occurs in the
bremsstrahlung processes. Because of the vanishing photon mass, the
fermion propagators become infinite when a quasi forward-scattered
quark emits a collinear photon: the singularities are regularized by an
interplay between the thermal masses of the fermion and the gluon. It
was seen before that this leads to an enhancement by a factor $1/g^2$ of
the bremsstrahlung contributions to the real soft photon rate so that
the two-loop diagrams entirely dominate over the one-loop contribution.
On the other hand, the bremsstrahlung production of a hard real photon
occurs at the same order in $g$ as the one-loop result. A rather simple
analytic expression, valid for soft, hard and ultra-hard photon has been
derived.  For ultra-hard photons, another process becomes dominant,
consisting of $q\bar{q}$ annihilation where one of the fermions
undergoes a scattering in the medium. In all cases, the calculated
two-loop contributions considerably increase the rate of photon or
lepton pair emission. Our results can easily be numerically extended to
cover, on the one hand, the case of a soft lepton pair at non-vanishing
momentum, and on the other hand, the case of lepton pairs produced at
large momentum with a small invariant mass. All the features of the
present results should survive in these more general situations.

A word of caution should be given concerning the bremsstrahlung
contribution to the soft virtual photon rate. The result shows a
sensitivity to a hard space-like gluon and therefore one may suspect that the
extrapolation of the effective gluon propagator in the hard region is
not complete: to be consistent may require  taking into account three-loop 
diagrams in very much the same way two-loop diagrams were needed
besides the one-loop diagram with a hard space-like fermion propagator.
On the contrary, for the case of a real photon emission, the 
bremsstrahlung rate
is sensitive only to soft gluon momenta, and the calculation is therefore
expected to be complete.

Our study does not cover all the physical processes included in the
two-loop diagrams. In particular, the  contribution with a time-like
gluon should be added. However such processes when the gluon is hard are
already contained in the one-loop diagram: counterterms should therefore
be included. To do this will be a very interesting practical exercise in
the use of  the effective theory up to two-loop.


\section*{Acknowledgments}
We thank R.~Baier for useful discussions. 
The work of RK was supported by the Natural Sciences and Engineering
Research Council of Canada. We also acknowledge support by NATO under grant CRG.
930739.
\appendix

\section{Hard thermal loops and counterterms}
\label{app:counterterms}
When using effective theories based on the summation of hard thermal loops at
higher orders, there is potentially a possibility to have multiple counting of
thermal corrections that should be there only once.

As a first illustration of this problem in a trivial context, let us use the
example of a massless real scalar field with a $\lambda^2\phi^4$
interaction in $4$ dimensions. The Lagrangian of such a model is
\begin{equation}
  {\cal L}\equiv {1\over
    2}\partial_\mu\phi\partial^\mu\phi-{{\lambda^2}\over{4!}}\phi^4\; .
\end{equation}
After the calculation of the one-loop tadpole, one realizes that this diagram
-- a hard thermal loop in the terminology of \cite{BraatP1} --
generates a thermal mass $m=\lambda T\surd 24$ that can be important for the
phenomenology of soft modes. Therefore, the idea of the HTL resummation is to
include this thermal mass in an effective Lagrangian
\begin{equation}
  {\cal L}_{\rm eff}\equiv {1\over 2}\partial_\mu\phi\partial^\mu\phi -{1\over
    2}m^2\phi^2 -{{\lambda^2}\over{4!}}\phi^4\; ,
\end{equation}
which is the Lagrangian of a real scalar field of mass $m$.

Let us assume now that one uses this effective theory instead of the bare one to
calculate the same tadpole diagram. The result would be
\begin{equation}
  \Pi={{\lambda^2T^2}\over{24}}\left(1-{3\over\pi}{m\over
      T}-{3\over{2\pi^2}}\left({m\over T}\right)^2\ln\left({m\over T}\right)
    +\cdots\right)=m^2(1+{\cal O}(\lambda))\; ,
\end{equation}
and since we start now from a propagator with squared mass $m^2$, the
resummation of the self-energy $\Pi$ would lead us to a propagator with a
squared mass $2m^2(1+{\cal O}(\lambda))$, approximately twice larger than the
correct thermal mass.
Obviously, the above result arises due to multiple 
countings of the same thermal
correction. Stated differently, this is a consequence of the fact that this
effective theory is more than a mere reordering of the 
perturbative expansion of
the bare theory since its Lagrangian is different. To solve this problem, one
must write
\begin{equation}
  {\cal L}={\cal L}_{\rm eff}+{\cal L}_{\rm ct} {\ \ \rm with\ \ }
  {\cal L}_{\rm ct}\equiv{1\over 2}m^2\phi^2\; ,
\end{equation}
and treat the counterterm ${\cal L}_{\rm ct}$ as an interaction term, just like
$\lambda^2\phi^4/4!$. The effect of this counterterm is of course to subtract
at higher order the thermal corrections that had already been included at the
tree level via the effective Lagrangian, in order to avoid multiple countings.
For instance, in the above example, the correct answer for the tadpole $\Pi$
when one takes care of the counterterm is
\begin{equation}
  \Pi={{\lambda^2T^2}\over{24}}\left(1-{3\over\pi}{m\over
      T}-{3\over{2\pi^2}}\left({m\over T}\right)^2\ln\left({m\over T}\right)
    +\cdots\right)-m^2={\cal O}(\lambda m^2)
\end{equation}
which is a perturbative correction to the thermal mass found at the previous
step, as it should be.
When this counterterm is taken properly into account, the effective theory is
nothing more than a reordering of the perturbative expansion, since the overall
Lagrangian remains unmodified.

\begin{figure}[htbp]
  \centerline{
    \resizebox*{!}{2.3cm}{\includegraphics{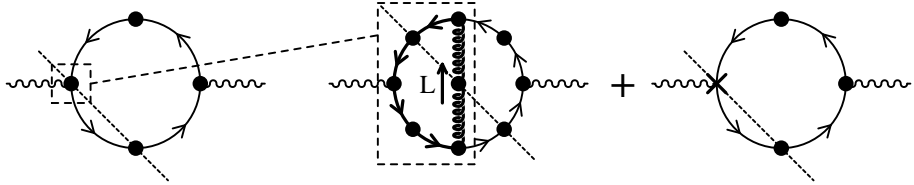}}
    }
  \caption{\footnotesize{Some corrections to the one-loop contribution (left
      diagram) to
      the photon polarization tensor. A black dot denotes an effective
      propagator or vertex. Crosses are HTL counterterms. The boldface 
      loop may
      reproduce what is already included in the one-loop diagram when its
      momentum is hard.}}
  \label{fig:counterterms}
\end{figure}

This problem also arises in effective gauge theories, where the situation is a
bit more complicated since we need there an infinite series -- one for each 
hard thermal loop -- of non-local counterterms. These counterterms are 
defined to
be the opposite of the HTL contribution to the corresponding function. Then, to
a given diagram obtained in the effective theory, one should add the diagrams
obtained by collapsing each loop in turn and replacing it by the corresponding
counterterm. In order to be more definite, let us consider the example of the
two-loop diagram of figure \ref{fig:2loop}-[b].
In the figure \ref{fig:counterterms}, we have represented next to the one-loop
contribution to the polarization tensor of the photon some of its two-loop
corrections.
It should be clear from the figure that when the boldface loop of the second
diagram carries a hard momentum, then this loop reproduces the HTL part of the
effective vertex in the dotted box already included at one-loop, plus new
sub-dominant perturbative corrections. The purpose of the third diagram is
precisely to subtract a quantity equal to the HTL contribution to this vertex,
so that what remains constitutes only new contributions. 
The net effect of this procedure
is thus to reorder the terms of the perturbative expansion.

Let us now explain why, despite their conceptual importance, the coun\-ter\-terms
do not contribute in the case of bremsstrahlung. As said in section
\ref{sec:bremss}, the contributions of bremsstrahlung to photon production come
from two-loop diagrams in which the gluon propagator is cut, and where one retains
only the Landau damping part ($L^2<0$) of the cut gluon propagator. 
Technically,
the gluon included in the 3-point HTL that appears
on the third diagram of Figure \ref{fig:counterterms} is a bare one, and
therefore its discontinuity has support totally included in the
time-like region. As a consequence, the diagram with counterterms contributes
only in the region of phase-space where $L^2\ge 0$, and cannot contribute to
bremsstrahlung. Even if they are not a worry in the case of bremsstrahlung, 
it was important to discuss the potential effect of
counterterms since, having shown that the second diagram gives an important
contribution, it is not enough to conclude that the two-loop order is important, 
as it may be canceled by counterterms.

\section{Phase space considerations}    
\label{app:phasespace}
The purpose of this appendix is to emphasize the importance of being in a
thermal bath in order to have a hard phase space for the quark loop. This is
indeed crucial in the calculations performed in the previous sections 
since this feature of the diagrams considered in this paper enables them to be
of the same order of magnitude as one-loop diagrams. In order to make the
following discussion more intuitive, let us first transform the bremsstrahlung
photon production rate in a way that separates more clearly the phase space from
the amplitude of the process producing the photon. The tools to do that have
already been presented in \cite{AurenGKP2} (see the section devoted to the
comparison with semi-classical methods), so that we only give the result
here\footnote{This formula is given here for the production of real photons by
bremsstrahlung. For other processes, analogue formulae still hold,
in which the amplitude squared have to be appropriately modified.}:
\setbox1=\hbox to 5cm{{\resizebox*{5cm}{!}{\includegraphics{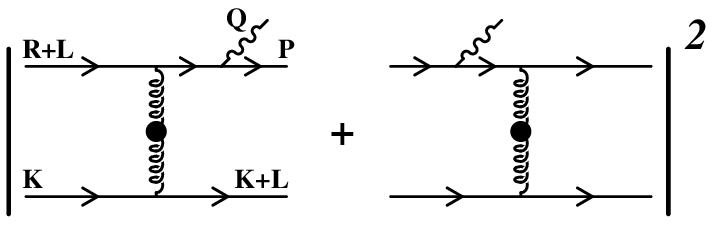}}}}
\begin{Eqnarray}
  &&\left.{{dN}\over{dtd{\imb x}}}\right|_{\rm bremss}={{d{\imb
        q}}\over{(2\pi)^32q_o}}\int{{d^4P}\over{(2\pi)^4}}
  \int{{d^4K}\over{(2\pi)^4}}
  \int{{d^4L}\over{(2\pi)^4}}\nonumber\\
  &&\qquad\times\raise -7mm\box1\nonumber\\
  &&\qquad\times2\pi\delta(P^2-M^2_\infty)\,
  2\pi\delta((R+L)^2-M^2_\infty)\nonumber\\
  &&\qquad\times2\pi\delta(K^2-M^2_\infty)\,
  2\pi\delta((K+L)^2-M^2_\infty)\nonumber\\
  &&\qquad\times n_{_{F}}(r_o+l_o)n_{_{F}}(k_o)
  [1-n_{_{F}}(p_o)][1-n_{_{F}}(k_o+l_o)]\; ,
  \label{eq:lastform}
\end{Eqnarray}
where $R\equiv P+Q$. 
This formula tells us that in order to calculate the contribution of some
process to the photon production
rate, we just have to integrate the amplitude squared of this
process over the phase space of unobserved particles (here, the incoming and
outgoing quarks). When doing this
integration, the external particles are put on their mass shells, and
accompanied by the appropriate statistical weight. The main advantage of this
formula is that it exhibits a clear separation in two factors: the amplitude
squared of the process producing the photon, and the phase space of the quarks.
Therefore, this formula can be used to separate the estimation of the order of
magnitude of the diagram in two steps: the order of magnitude of the amplitude,
and the size of the phase space over which it must be integrated.

Looking at this formula, it is clear that the effect of the
thermal bath appears only in the phase space. Indeed, the amplitude that appears
in Eq.~(\ref{eq:lastform}) is nothing but a zero temperature one (it does not
contain any statistical weight). Stated differently, if, instead of looking at
photon production by a thermal bath, we were looking at bremsstrahlung photon
production in p-p collisions ({\it i.e.} the two scattering quarks come from
protons in colliding beams), the statistical factors of incoming quarks would have
to be replaced\footnote{To understand the analogy between the two situations,
one may see a proton beam as a dense medium containing quarks and gluons, with
distributions related to the structure functions of quarks and gluons inside the
proton. Such a medium is highly anisotropic, since all the partons go in the
direction of the beam.} by structure functions\footnote{More exactly, 
to obtain the
correctly normalized photon production rate, one should also multiply by 
the proton densities in
the beams.} of a quark inside a proton, and the
statistical weights of outgoing quarks would be replaced by 1. 
This difference is precisely
the point which makes the thermal bath dramatically different from 
the p-p collision.
Indeed, in the case of the thermal bath, the statistical functions have a
support which is 3-dimensional since the plasma is an isotropic medium. On the
other hand, the structure function of the quark inside protons of the beam is
vanishing if the quark has a direction different from that of the beam. This
will make a difference when one performs the integral over the momentum $p$.
Indeed, if one has something like $\int dp p^n f(p)$ in the case of the thermal
bath, the integral would be $\int dp p^{n-2} f(p)$ for the case of p-p
collisions. As a consequence, the integral over the quark momentum is more
likely to be sensitive to hard momenta in the case of the thermal bath.
Therefore, even if it is not a rigorous proof, these considerations show why
some higher order 
processes which are not dominant at zero temperature may become dominant in
a plasma, due to a bigger size for their phase space.

Moreover, Eq.~(\ref{eq:lastform}) may help to understand why two-loop
contributions may be as important as one-loop ones. Indeed, it shows that the
final order of magnitude of a contribution results from a competition between
two effects. The first one is the order of magnitude of the amplitude, which
usually becomes smaller when the number of loops increases since the number of
coupling constants increases also. The second aspect of the problem is the size
of the phase space, since it may happen that due to kinematical constraints, the
one-loop phase space is much smaller than the two-loop one. Both effects can
compensate so that two-loop diagrams contribute also at the dominant level. 
This is
precisely what happens in the case of photon production by a plasma: the one-loop
phase space is soft due to kinematical constraints, while the two-loop thermal phase
space can be hard (as explained above, this is possible because we are in an
isotropic medium). Therefore, the smallness of the two-loop amplitude is
compensated by the size of the two-loop phase space.

\end{document}